# Click-Through Rate Prediction in Online Advertising: A Literature Review

Yanwu Yang[1], Panyu Zhai[1]
[1]School of Management, Huazhong University of Science and Technology, Wuhan, China
{yangyanwu.isec,zhaipanyu.isec}@gmail.com

**Abstract**: Predicting the probability that a user will click on a specific advertisement has been a prevalent issue in online advertising, attracting much research attention in the past decades. As a hot research frontier driven by industrial needs, recent years have witnessed more and more novel learning models employed to improve advertising CTR prediction. Although extant research provides necessary details on algorithmic design for addressing a variety of specific problems in advertising CTR prediction, the methodological evolution and connections between modeling frameworks are precluded. However, to the best of our knowledge, there are few comprehensive surveys on this topic. We make a systematic literature review on state-of-the-art and latest CTR prediction research, with a special focus on modeling frameworks. Specifically, we give a classification of state-of-the-art CTR prediction models in the extant literature, within which basic modeling frameworks and their extensions, advantages and disadvantages, and performance assessment for CTR prediction are presented. Moreover, we summarize CTR prediction models with respect to the complexity and the order of feature interactions, and performance comparisons on various datasets. Furthermore, we identify current research trends, main challenges and potential future directions worthy of further explorations. This review is expected to provide fundamental knowledge and efficient entry points for IS and marketing scholars who want to engage in this area.

# 1. Introduction

The Internet has promised a variety of online advertising forms leveraging many digital media vehicles (e.g., search portals, social media platforms, e-commerce platforms, online games, mobile apps, online videos, banners, etc.) to deliver marketing messages to potential consumers (Yang et al., 2017). Online advertising has become a dominant sector in the advertising industry. According to the Statista report (Statista, 2021), online advertising revenue in the United States grew by 12.2 percent in 2020 compared to 2019, from $124.6 billion to $139.8 billion. The online advertising market is expected to reach $982.82 billion by 2025 (Mordor Intelligence, 2021).

In online advertising, click-based performance indexes, e.g., clicks and click-through rate (CTR) reflect the relevance of advertisements from users' perspective. As recognized by both researchers and practitioners, improving CTR is an effective way to realize the sustainable development of online advertising ecosystems (Robinson et al., 2007; Rosales et al., 2012; Tan et al., 2020). Hence, advertising CTR prediction has attracted much research attention in the past decades (Yan et al., 2014; Chapelle et al., 2014; McMahan et al., 2013; Richardson et al., 2007).

As a hot research frontier driven by industrial needs, recent years have witnessed more and more novel learning models employed to improve advertising CTR prediction. Although extant research provides sufficient details on algorithmic design for addressing a variety of specific problems related to this topic, the methodological evolution and connections between modeling frameworks are precluded. However, to the best of our knowledge, there are few comprehensive surveys on CTR prediction in the context of online advertising.

The objectives of our review are two-fold. First, we aim to make a systematic literature review on existing CTR prediction research, with a special focus on modeling frameworks. Second, we identify current research trends, main challenges and potential future directions worthy of further explorations.

This review complements review articles recently published on users' responses (including CTR, conversion rate and user engagement) prediction (Gharibshah and Zhu, 2021) and CTR prediction (Wang, 2020; Zhang et al., 2021a). More specifically, Gharibshah and Zhu (2021) focused on online advertising platforms, data sources and features, and typical methods for user response prediction; Wang (2020) briefly introduced several classical methods for CTR prediction; and Zhang et al. (2021a) concentrated on the transfer from shallow to deep learning models for CTR prediction, explicit feature interaction modules and automated methods for architecture design.

Our review is different from existing ones in the following aspects. First of all, we provide a comprehensive review on CTR prediction by organizing the development of extant research in a synthetic framework emphasizing connections between CTR prediction models. In particular, we give preliminaries on the problem of advertising CTR prediction, present state-of-the-art models, and outline major challenges and research perspectives in this area. For each category of CTR prediction models, we introduce the basic framework, its variants and ensemble models, and discuss advantages and disadvantages, and performance evaluation on various datasets. Second, this review focuses on CTR prediction models in the context of online advertising, excluding those in other contexts such as recommender system and Web search. Although Web search, recommender system and online advertising are three popular information-seeking mechanisms to mitigate the information overload problem (Zhao et al., 2019), they are differ with respect to inputs, outputs and goals (see Appendix A.1). Third, this review covers existing research on CTR prediction almost without reservation, by searching over six major academic databases. In other words, this review contains a complete map of state-of-the-art and latest models proposed for advertising CTR prediction.

The rest of this paper is organized as follows. Section 2 describes the search procedure for identifying articles covered in this review. Section 3 gives preliminary knowledge on the topic of advertising CTR prediction, including problem definition, foundational concepts, procedure, features and evaluation metrics. Section 4 gives a classification of state-of-the-art CTR prediction models in the extant literature and presents modeling frameworks, advantages and disadvantages, and performance assessment for CTR prediction. Section 5 summarizes CTR prediction models with respect to the complexity and the order of feature interactions, and performance evaluation on various datasets. Section 6 discusses current research trends, main challenges and future directions in the domain of advertising CTR prediction. Section 7 concludes this review.

## 2. Literature Search and Study Identification

This review covers publications mainly from six academic databases: Web of Science, ACM, IEEE, EBSCOhost, ScienceDirect and ABI/Inform Global using full text search of keywords ("click through rate prediction" OR "CTR prediction" OR "click prediction") AND ("online advertising" OR "internet advertising" OR "digital advertising" OR "social media advertising" OR "display advertising" OR "mobile advertising" OR "social network advertising"). This process led to 101 results from Web of Science, 246 results from ACM, 106 results from IEEE, 138 results from EBSCOhost, 30 results from ScienceDirect, 47 results from ABI/Inform Global. We eliminated duplicate articles, dropped patents, reviews, news, magazines, reports, posters, and abstracts, which yielded 468 articles. Moreover, in order to get a broader coverage on publications on this topic, we retrieved citations and related articles of those identified in the previous step in Google scholar, which identified additional 20 results. Manual screening was made to identify whether each article addresses advertising CTR prediction problems by going through its title, abstract, full-text and datasets, which excluded 356 articles. Note that studies conducting CTR prediction experiments on advertising datasets (e.g., Criteo, Avazu and KDDCUP 2012 track 2) were counted in this review. Finally, this process resulted in 132 articles, including 26 peer-reviewed journal articles, 93 conference articles, 13 pre-prints. Our search and selection process is illustrated in Figure 1. Table A1 in the appendix gives publication date, authors, context and publication venue of articles included in this review.

Figure 2 visualizes the distribution of selected studies for this review on publication year. As we can see, publications on CTR prediction have been increasing in an exponential progression since 2007, reached a peak period in the past five years (i.e., 2016-2020). This indicates that it is an active topic which has attracted plentiful research efforts. Moreover, the increasing trend shows that more studies on CTR prediction will be reported in the future.

## 3. Preliminaries

This section first presents preliminary knowledge for CTR prediction in online advertising, including problem definition, features and evaluation metrics commonly used in the literature. For details about the general procedure of CTR prediction, see Appendix A.2.

### 3.1 The Definition of CTR Prediction Problem

In online advertising, the CTR prediction problem can be defined as follows. Given a set of triples $\langle u_i, x_i, y_i \rangle, i = 1,2, \dots, N$ where $u_i$ and $x_i$ represents user and advertisement, respectively, in the $i-th$ instance, and $y_i = 1$ means that $u_i$ clicks on $x_i$, otherwise $y_i = 0$, the CTR prediction problem is to predict whether user $u_j$ will click on advertisement $x_j$ in the $j-th$ instance, i.e., $p = f(u_j, x_j)$ where $p$ is termed as the click-through rate (CTR).

Mathematically, a function $p = f(u, x)$ can be formulated to derive CTR by minimizing a loss function of cross entropy (Zhang et al., 2016; Li et al., 2020a), which is given as follows.

$$L = -\frac{1}{N}\sum_{i=1}^{N}[y_i \times logp_i + (1 - y_i)\,(1 - logp_i)], \qquad (1)$$

where $y_i$ = 0 (or 1) denotes the target label, $p_i \in [0,1]$ is the predicted CTR.

Notations used in this paper are presented in Table 1.

## 3.2 Features for CTR Prediction

In the literature on advertising CTR prediction (e.g., Shan et al., 2016; Deng et al., 2018; Ling et al., 2017; Liao et al., 2014; Li et al., 2015; Zhang et al., 2017), features can be categorized into five classes, as summarized in Table 2: (1) advertising features; (2) user features; (3) context features; (4) query features; and (5) publisher features.

Feature engineering is mostly needed for transforming raw data into useful features for CTR prediction, either manually or automatically. In traditional CTR prediction models (e.g. logistic regression), manual feature engineering requires massive time and relies on experts' experiences. Recently, the literature on prediction models enabling automatic feature engineering (e.g., PNN, DeepFM) has started to emerge.

The commonly used feature engineering techniques include one-hot encoder, data standardization, relationship analysis between features and labels, and feature combinations (Ren et al., 2020). Counting features are a recently developed statistics-based feature engineering technique to address cold-start and data drift problems (Wu et al., 2019). For details on feature engineering in machine learning, refer to Dong & Liu (2018) and Kuhn & Johnson (2019).

## 3.3 Evaluation Metrics for CTR Prediction

In the literature, a set of metrics has been introduced and commonly employed to evaluate the performance of CTR prediction models, including precision, recall, F1-score, accuracy, area under the curve of receiver operating characteristic (AUC-ROC), area under the curve of precision/recall (AUC-PR), relative improvement (RelaImpr), Logloss, mean square error (MSE), root means squared error (RMSE), relative information gain (RIG) and field-level calibration errors. Table 3 illustrates definitions of evaluation metrics and related studies. For more details on evaluation metrics, refer to Ricci et al. (2015).

# 4. State-of-the-art CTR Prediction Models

State-of-the-art CTR prediction models reported in the advertising literature can be categorized into four groups: (1) Multivariate statistical models; (2) Factorization machines (FMs) based models; (3) Deep learning models; (4) Tree models. Table 4 presents categories of state-of-the-art CTR prediction models and the distribution of selected studies in our review. In the following, for each category of models, we will go over its modeling frameworks, advantages and disadvantages, and performance assessment for CTR prediction.

## 4.1 Multivariate Statistical Models

In the literature on advertising CTR prediction (e.g., Yan et al., 2014; Richardson et al., 2007; Ling et al., 2017; Chang et al., 2010; Juan et al., 2016), two multivariate statistical methods, namely logistic regression and degree-2 polynomial, have been employed to explore various factors affecting users' response behaviors toward advertising (e.g., clicks).

### 4.1.1 Logistic Regression (LR)

Logistic regression (LR) is one of the most widely used methods in multi-feature CTR prediction problems (Kumar et al., 2015). Figure 3 gives the LR modeling framework. Typically, LR is given as follows.

$$\Phi_{LR}(\boldsymbol{w},\boldsymbol{x}) = \boldsymbol{w}^T\boldsymbol{x} + b, \qquad (2)$$

$$p = \sigma\big(\Phi_{LR}(\boldsymbol{w},\boldsymbol{x})\big), \qquad (3)$$

where $\Phi_{LR}(\boldsymbol{w},\boldsymbol{x})$ is a linear combination of individual features (Pan et al., 2018), $p$ is the predicted value, $\sigma$ represents the activation function, $\boldsymbol{x} = [x_1, x_2, x_3, \dots, x_{n-1}, x_n]$ is the input vector, $\boldsymbol{w} = [w_1, w_2, w_3, \dots, w_{n-1}, w_n]$ is the weight vector to be estimated.

For each single instance, the loss is $l_i(w) = -[y_i \times logp_i + (1-y_i)(1-logp_i)]$, where $y_i$ and $p_i$ are the observed and predicted values for instance $i$, respectively. The gradient of loss can be represented as $\nabla l_i(w) = (p_i - y_i)x_i$. The min-batch gradient descent (MBGD) method can used to update the weight vector $w$ in min-batches (Goodfellow et al., 2016; Gou & Yu, 2018; Jie-Hao et al., 2017; Zhou et al., 2018; Jiang et al., 2017), i.e., $w = w - \alpha\frac{1}{I}\sum_{i=1}^{I}(p_i - y_i)\cdot x_i$, where $I$ is the number of batch instances, $\alpha$ is the learning rate.

The LR-based CTR prediction model has several advantages: (1) it can obtain a value between 0 and 1 describing the click probability; (2) it is a linear combination of various features, capturing correlations between features and the label; (3) its gradient of loss function has an elegant form (Richardson et al., 2007; Moneera et al., 2021). However, LR fails to represent interactive effects and non-linear relationships between features and the label because it assumes each feature to be independent (Ling et al., 2017).

Generally, in the literature (e.g., Guo et al., 2017; Wang et al., 2016; Pan et al., 2018), LR usually serves as the baseline in performance evaluation of proposed CTR prediction models. Moreover, LR can be combined with other methods to predict clicks, e.g., FMSDAELR (Jiang et al., 2018), FO-FTRL-DCN (Huang et al., 2020) and GBDT+LR (He et al., 2014), where it forms the top layer after feature selection and feature interactions. We will discuss in detail FMSDAELR and FO-FTRL-DCN in Section 4.3 and GBDT+LR in Section 4.4.

### 4.1.2 Degree-2 Polynomial (Poly2)

Linear models such as LR perform less well for CTR prediction because they do not consider possible relationships among features (i.e., feature interactions) (Chapelle et al., 2014; Pan et al., 2018). Feature interactions can generate new useful features that can significantly improve the performance of CTR prediction models. In most cases with linear CTR prediction models, features interactions are made manually. However, there is no commonly accepted operation procedure for manual feature interactions (Ren et al., 2020). That is, the effect of manual feature interactions, to a great degree, depends on experts' domain-specific knowledge and experience. Therefore, how to derive interactive relationships between features becomes a critical issue.

The Degree-2 Polynomial (Poly2) can explicitly describe interactions between a pair of features (Chang et al., 2010). Figure 4 gives the Poly2 modeling framework. Following Pan et al. (2018) and Jie-Hao et al. (2017), Poly2 can be formulated as:

$$\Phi_{Poly2}(w, x) = w_0 + \sum_{i=1}^{n} w_i x_i + \sum_{i=1}^{n} \sum_{j=i+1}^{n} w_{h(i,j)} x_i x_j, \qquad (4)$$

where $\Phi_{Poly2}$ is a linear sum of individual features ($x_i$) and combinatorial features ($x_i x_j$) (i.e., feature interactions), $w_i$ denotes the strength of a single feature, $w_0$ is the global bias, and $w_{h(i,j)}$ denotes the strength of combinatorial features. $h(i,j)$ is a function mapping $i$ and $j$ into a natural number in the hashing space so as to reduce the dimension, such as $h(i,j) = ni + j$ (Guo et al., 2018; Chang et al., 2020).

Compared to linear models, Poly2 performs better on CTR prediction in accuracy (Chang et al., 2010). However, Poly2 is slower because of its higher complexity. Moreover, in some cases, sparsity may remain even after performing feature interactions so that Poly2 performs less well on sparse data (Chang et al., 2020; Pan et al., 2018).

In addition to LR and Poly2, traditional classification models (e.g., linear regression, Poisson regression, decision tree and support vector machines) have also been applied to CTR prediction (Wang & Chen, 2011; Wang et al., 2012; 2013; Avila Clemenshia & Vijaya, 2016). However, these models are underperformed in the sparse problem because they fail to capture feature interactions which are essential for CTR prediction.

## 4.2 Factorization Machines (FMs) based Models

In order to solve the sparse data problem for CTR prediction, researchers have explored non-linear feature interactions in the framework of factorization machines (FMs). In the following we go through three major FM-based modeling frameworks, namely, the classic Factorization Machines (FMs), Field-aware Factorization Machines (FFMs) and Field-weighted Factorization Machines (FwFMs). Among the three FMs-based modeling frameworks, FMs emphasize feature interactions, while FFMs consider both feature interactions and field interactions, and FwFMs are an improved combination of FMs and FFMs (Gharibshah & Zhu, 2021; Juan et al., 2016; Juan et al., 2017; Pan et al., 2018; Zhang et al., 2019).

### 4.2.1 Factorization Machines (FMs)

Factorization Machines (FMs) can support better model estimation on sparse data by using factorized parameters. Figure 5 gives the FMs modeling framework. Typically, FMs are given as follows (Rendle, 2010; Wang et al., 2020c).

$$\Phi_{FMs}(\boldsymbol{w},\boldsymbol{v},\boldsymbol{x}) = w_0 + \sum_{i=1}^{n} w_i x_i + \sum_{i=1}^{n}\sum_{j=i+1}^{n}\langle \boldsymbol{v}_i,\boldsymbol{v}_j\rangle x_i x_j, \qquad (5)$$

where $\boldsymbol{v}_i \in \mathbb{R}^k$ ($\boldsymbol{v}_j \in \mathbb{R}^k$) are the embedding vectors of $x_i$ ($x_j$), which can be regarded as the contribution of $x_i$ ($x_j$) when it interacts with feature $x_j$ ($x_i$). $k$ is the dimensionality of the factorization; $w_0$ is the global bias, $w_i$ and $\langle \boldsymbol{v}_i,\boldsymbol{v}_j\rangle$ describe the strength of a single feature and that of a combinatorial feature (i.e., feature interactions), respectively; $\langle\cdot,\cdot\rangle$ is the dot product of two vectors of size $k$. $\boldsymbol{v}_i$ and $\boldsymbol{v}_j$ are the vectors of parameters to be estimated. The dot product of $\boldsymbol{v}_i$ and $\boldsymbol{v}_j$ is given as $\langle \boldsymbol{v}_i,\boldsymbol{v}_j\rangle = \sum_{f=1}^{k} v_{if}\cdot v_{jf}$, where $v_{if}$ and $v_{jf}$ are the $f-th$ dimensional component of $\boldsymbol{v}_i$ and $\boldsymbol{v}_j$, respectively. FMs use the dot product of the two feature vectors to represent feature interactions.

Note that FMs differ from Poly2 in handling interaction terms (i.e., parameters of feature interactions): for each feature interaction, FMs factorize the associated parameter, while Poly2 simply assigns a parameter $w_{h(i,j)} \epsilon \mathbb{R}$. Parameters of feature interactions in Poly2 can be regarded as a $n \times n$ matrix, which is factorized into a product of two matrices in FMs.

Attributing to the parameter factorization, FMs perform better than Poly2 in RMSE in applications such as CTR prediction where the data is sparse. On one hand, FMs results in additional data for parameter estimation, by breaking the independence of interaction parameters (Rendle, 2010). An interesting effort by Pan et al. (2016) introduced the Laplace distribution into FMs to further fit the sparse data. On the other hand, in FMs, the data for a feature interaction can also support estimation of related interactions. That is, in the case that a feature pair does not appear sufficient times, interaction parameters of FMs can be learned from related feature interactions, while the estimation of Poly2 may be inaccurate. Moreover, FMs have a linear complexity in both $k$ and $n$ (Jie-Hao et al., 2017), i.e., the number of parameters to be estimated is linearly proportional to the dimensionality of the factorization and the number of features (Pan et al., 2018). For a summary of FMs learning and optimization algorithms, refer to see Rendle (2012a).

In the literature on advertising research, FMs and their extensions have been widely employed for CTR prediction. In the KDDCup 2012 competition, Rendle (2012b) showed that the FMs can estimate feature interactions reliably in predicting advertising CTR. Ta (2015) proposed an online learning algorithm for CTR prediction incorporating the Follow-The-Regularized-Leader (FTRL-proximal) algorithm with per-coordinate learning rate into FMs, which outperforms FMs with the stochastic gradient method (SGD) in AUC. For sparse settings, He & Chua (2017) proposed neural factorization machine (NFM) by exploiting high-order feature interactions. NFM utilizes the bi-interaction pooling to describe second-order feature interactions, and then uses a deep neural network to capture high-order feature interactions.

In order to characterize the importance of various feature interactions, Xiao et al. (2017) proposed the attentional factorization machine (AFM) which allows different feature interactions to contribute differently based on an attention-based pooling mechanism. AFM uses the element-wise product of two vectors to encode feature interactions on dense vectors. Compared to deep learning models (e.g., Wide & Deep, DeepCross), NFM and AFM can obtain better performance for CTR prediction, but with a shallower structure and with fewer parameters, respectively. For explicitly representing high-order feature interactions, Tao et al. (2020) developed High-order Attentive Factorization Machine (HoAFM) where a cross interaction layer updates feature representation based on co-occurrence and a bit-wise attention mechanism determines importance of co-occurred features, which outperforms FMs and NFM in AUC and Logloss on benchmark advertising datasets.

#### 4.2.2 Field-aware Factorization Machines (FFMs)

Advertising CTR prediction usually involves massive multi-field categorical data (e.g., gender and age). Although FMs perform well in big sparse data, they fail to capture the fact that a feature might behave differently when interacting with various features from multiple fields. To this end, Juan et al. (2017) introduced Field-aware Factorization Machines (FFMs) and developed the formulation of interactions. Figure 6 presents the FFMs modeling framework. Following Juan et al. (2017), Pan et al. (2018) and Zhang et al. (2019), FFMs are given as

$$\Phi_{FFMs}(\boldsymbol{w}, \boldsymbol{v}, \boldsymbol{x}) = w_0 + \sum_{i=1}^{n} w_i x_i + \sum_{i=1}^{n} \sum_{j=i+1}^{n} \langle \boldsymbol{v}_{i,F(j)}, \boldsymbol{v}_{j,F(i)} \rangle x_i x_j, \qquad (6)$$

where $F(i)$ and $F(j)$ represent fields of $x_i$ and $x_j$, respectively, $\boldsymbol{v}_{i,F_{(j)}} \epsilon \mathbb{R}^k$ ($\boldsymbol{v}_{j,F_{(i)}} \epsilon \mathbb{R}^k$) is the interaction strength when $x_i$ ($x_j$) interacts with $x_j$ ($x_i$); other modeling parameters are identical to those in FMs.

Compared to FMs, FFMs increase the number of parameters in each feature interaction (Phangtriastu & Isa, 2018). In FFMs, when a categorical feature (e.g., Female) interact with features in other fields (e.g., country and age), interaction parameters $\boldsymbol{v}_{Female,Country}$ and $\boldsymbol{v}_{Female,Age}$ should be different from each other. Juan et al. (2016) evaluated LR, Poly2, FMs, and FFMs with the stochastic gradient descent (SGD) method in advertising CTR prediction, and demonstrated that FFMs can outperform LM, Poly2 and FMs in Logloss, while at the cost of longer training time. In order to overcome the limitation of computing on a single machine, Ma et al. (2016) implemented a distributed architecture of FFMs.

Different operations such as convolutional, inner-product, out-product play different roles in feature interactions. In order to handle feature interactions more effectively, Yang et al. (2020) proposed a neural model named operation-aware neural networks (ONNs) using an operation-aware embedding method to learn feature representations for different operations, which can outperform FFMs for CTR prediction tasks in AUC, Logloss and RMSE. Conceptually, the operation-aware method can be considered as a generalization of the field-aware method because in the former both a same type of operation performing on different features and different types of operations on a same feature are regarded as different operations. Similar to FFMs, ONNs use an embedding vector for second-order feature interactions, and the output of the interactions, together with the embedding vector, serve as the input of the multi-layer perceptron (MLP) (Samel, 2017) to make prediction.

In order to achieve a high-order feature learning ability, Zhang et al. (2019) proposed the field-aware neural factorization machine (FNFM) incorporating a deep neural network. FNFM performs the field-aware second-order feature interaction learning with a bi-interaction layer. Essentially, FNFM is an end-to-end modeling framework combining FFMs and DNN.

With consideration of the fact that feature interactions may have different importance for CTR prediction, Wang et al. (2020c) proposed an attention-over-attention field-aware factorization machine (AOAFFM) which employs a feature-level attention and an interaction-level attention to distinguish importance of feature interactions, which substantially improves FMs and FFMs.

### 4.2.3 Field-weighted Factorization Machines (FwFMs)

In advertising CTR prediction, a large number of features significantly increase the modeling complexity for FFMs. That is, the number of parameters in FFMs increases with the order of feature number times field number. In order to reduce parameters to be estimated, Pan et al. (2018) extended FFMs by treating interaction terms and field pairs separately with a weighting mechanism, called field-weighted factorization machines (FwFMs). FwFMs are given as:

$$\Phi_{FwFMs}(\boldsymbol{w}, \boldsymbol{v}, \boldsymbol{x}) = w_0 + \sum_{i=1}^{n} w_i x_i + \sum_{i=1}^{n} \sum_{j=i+1}^{n} \langle \boldsymbol{v}_i, \boldsymbol{v}_j \rangle r_{F(i),F(j)} x_i x_j, \qquad (7)$$

where $r_{F(i),F(j)} \in \mathbb{R}$ denotes the strength of interaction between two fields $F(i)$ and $F(j)$; other modeling parameters are identical to those in FMs and FFMs. Note that FwFMs use the interaction weight $r_{F(i),F(j)}$ to explicitly describe the interaction strength for each pair of fields, while FFMs do it in an implicit way, i.e., by using the embedding vector $\boldsymbol{v}_{i,F(j)}$.

FwFMs can outperform LR, Poly2 and FMs, and achieve comparable performance with FFMs in AUC, while with fewer parameters to be estimated (Pan et al., 2018). In order to reduce the dimension of sparse data, Zou et al. (2020a) proposed a factorized weight interaction neural network (INN) method by mining high-order feature latent patterns, which performs better than NFM and AFM in RMSE and Logloss. Essentially, INN is similar to FwFMs in that it uses a weighted-interaction layer to represent feature interactions.

Recently, Sun et al. (2021) proposed field-matrixed factorization machines (FmFMs) for CTR prediction by representing feature interactions between field pairs as a matrix, which can outperform LR, FMs, FFMs and FwFMs, and have comparable performance to complex DNN models in AUC and Logloss. FmFMs can be considered as a generalization of FMs and FwFMs.

## 4.3 Deep Learning Models

We present modeling frameworks with 1-order and/or 2-order feature interactions and their performance in advertising CTR prediction in Sections 4.1 and 4.2. Although low-order models are easy to implement, they suffer from the underperformed prediction ability (Guo et al., 2018; Lian et al., 2018; Yan et al., 2020; Lian & Ge, 2020), especially in complex advertising systems (Yang et al., 2018). Hence, it calls for prediction models enabling the high-order feature representation. In this section, several major deep learning modeling frameworks for advertising CTR prediction with the high-order feature representation are discussed, including standard long short-term memory (LSTM) (Hochreiter and Schmidhuber, 1997), convolutional neural network (CNN) (Liu et al., 2015; Chan et al., 2018; Emmert-Streib et al., 2020), factorization machine supported neural network (FNN) (Zhang et al., 2016) and DeepFM (Guo

et al., 2017; 2018).

### 4.3.1 Long Short-Term Memory (LSTM)

The standard LSTM recurrent network cell is composed of three gates: input gate $\boldsymbol{i_t}$, forget gate $\boldsymbol{f_t}$, output gate $\boldsymbol{o_t}$. Note that gates are a distinctive feature of LSTM from recurrent neural networks (RNN): rather than using a simple RNN unit to deal with the weighted sum of input sequences, LSTM takes a linear self-loop cell memory with three different gates. The three gates are with sigmoid activation functions to control the amount of information passing through the cell memory: all the information flow when gate values are 1, and no information flow when the values are 0.

The basic principle of the standard LSTM network can be mathematically represented as the following system of equations (Wei et al., 2021).

$$\begin{cases} \boldsymbol{i_t} = \sigma(\boldsymbol{W_i x_t} + \boldsymbol{U_i h_{t-1}} + \boldsymbol{b_i}) \\ \boldsymbol{f_t} = \sigma\left(\boldsymbol{W_f x_t} + \boldsymbol{U_f h_{t-1}} + \boldsymbol{b_f}\right) \\ \boldsymbol{o_t} = \sigma(\boldsymbol{W_o x_t} + \boldsymbol{U_o h_{t-1}} + \boldsymbol{b_o}) \\ \boldsymbol{g_t} = tanh(\boldsymbol{W_c x_t} + \boldsymbol{U_c h_{t-1}} + \boldsymbol{b_c}) \\ \boldsymbol{c_t} = \boldsymbol{f_t} * \boldsymbol{c_{t-1}} + \boldsymbol{i_t} * \boldsymbol{g_t} \\ \boldsymbol{h_t} = \boldsymbol{o_t} * \boldsymbol{g_t} \end{cases}, \quad (8)$$

where $\boldsymbol{W_i}$, $\boldsymbol{W_f}$, $\boldsymbol{W_o}$, $\boldsymbol{W_c}$ denote input weight matrices in the three gates and the cell state, respectively; $\boldsymbol{U_i}$, $\boldsymbol{U_f}$, $\boldsymbol{U_o}$, $\boldsymbol{U_c}$ denote interconnected weight matrices in the three gates and the cell state, respectively; and $\boldsymbol{b_i}$, $\boldsymbol{b_f}$, $\boldsymbol{b_o}$, $\boldsymbol{b_c}$ are the corresponding bias terms.

Figure 7 presents the learning process of the LSTM network. At each step $t$, first of all, the input gate $\boldsymbol{i_t}$ controls the input flowing of information into the cell that receives the current input information $\boldsymbol{x_t}$ and the previous hidden state information $\boldsymbol{h_{t-1}}$, and the input information is squashed into features through a $tanh(\cdot)$ function in $\boldsymbol{g_t}$ (Pham et al., 2017). Second, the forget gate $\boldsymbol{f_t}$ determines how much previous information should be retained or forgotten from the cell as the storage of historical information, and the memory cell $\boldsymbol{c_t}$ is updated through the adjusted previous memory cell $\boldsymbol{c_{t-1}}$ with the forgetting gate output $\boldsymbol{f_t}$ and the activated input feature $\boldsymbol{i_t}$ with the difference between current and previous memory states. Third, the output gate $\boldsymbol{o_t}$ controls the output flowing of information from the memory cell (Chung et al., 2014), and the current hidden state $\boldsymbol{h_t}$ is updated based on $\boldsymbol{g_t}$ and the output gate $\boldsymbol{o_t}$.

The LSTM network initiatively proposed by Hochreiter and Schmidhuber (1997) is an improvement of RNN in that LSTM can overcome the shortcoming of RNN when the data has long-term dependencies (Lipton et al., 2015; Chen et al., 2016b; Emmert-Streib et al., 2020). Moreover, LSTM can effectively solve the gradient vanishing and exploding problem when using SGD to find the optimal solution (Hochreiter, 1998; Gers et al., 1999; Chen et al., 2016a).

Taking user's response behaviors toward advertisements displayed to her as temporal events, LSTM and RNN have been applied as the basic framework to predict whether a user is going to click an advertisement or not. As reported in the literature (Zhang et al., 2014b; Gharibshah et al., 2020), LSTM-based and RNN-based CTR prediction models outperform linear models (e.g., LR and Naive Bayes) and nonlinear models (e.g., neural networks and random forest) in AUC. However, LSTM is trained in sequence, as shown in Figure 7, thus needs more time and larger memory in the training step.

In order to capture latent interest behind concrete advertising behaviors such as clicks, Zhou et al. (2019) proposed a CTR prediction model named deep interest evolution network (DIEN) using gated recurrent unit (GRU) to capture the dependency between sequential behaviors, which can obtain a comparable performance with LSTM but with a lower computational complexity. GRU is similar to LSTM in the basic recurrent neural network architecture (Pi et al., 2019; Zhang et al., 2020). In related research, bidirectional GRU and LSTM have been combined with stack autoencoder to capture latent interests behind users' behaviors (Wang et al., 2019; 2020a).

The number of clicks on an advertisement may be affected by other advertisements displayed together with it on a Web page, which is called advertising externalities (Xu et al., 2010; Xiong et al., 2012). Deng et al. (2018) proposed a LSTM-RNN architecture to predict the click probability for a given advertisement on different positions, while considering of advertisements ranking higher than it, which performs better than DNN without consideration of the externality effect in AUC and RIG.

### 4.3.2 Convolutional Neural Network (CNN)

Convolutional neural network (CNN) is a type of feed-forward neural network composed with one or more convolutional layers, one or more pooling layers and a fully-connected layer, as illustrated in Figure 8.

Attributing to the design of shared weights mechanism, CNN is powerful to find local feature interactions and reduce the number of parameters (Lian and Ge, 2020). In CNN, an instance composed with many fields is embedded into dense vectors: each field is mapped into an embedding feature vector with a fixed length; the convolutional layer uses at least one kernel to slide across the instance matrix constructed by embedding feature vectors (Emmert-Streib et al., 2020); the pooling layer uses some specified pooling method to reduce the input dimension. An activation function is applied on the output of the pooling layer. The convolutional layer, pooling layer and activation function can achieve the multiple-order feature learning. All learned features are recombined by the fully-connected layer and the prediction can be obtained via the softmax function.

The convolutional click prediction model (CCPM) extracts local-global key features from an input instance with varied elements, which outperforms LR and FMs on single ad impression and RNN on sequential ad impression in Logloss (Liu et al., 2015). In CCPM a flexible $p$-max pooling layer is leveraged to select prominent features where $p$ is defined as a function of the number of convolutional layers and the length of the input instance. As convolution layers and pooling layers capture information in local respective fields, the sequence of embedding feature vectors may significantly affect feature interactions. However, in most practical applications such as online advertising, varying sequences of embedding feature vectors do not change the implied meaning. Moreover, useful feature interactions may be lost when only considering neighboring features (Chan et al., 2018; Liu et al., 2019).

Recently, researchers have made efforts to overcome shortcomings of CCPM by investigating improvements of CNN-based models for advertising CTR prediction, e.g., multi-sequence model (MSM) (Chan et al., 2018), feature generation by convolutional neural network (FGCNN) (Liu et al., 2019), and dense matrix based convolutional neural network (DMCNN) (Niu & Hou, 2020). MSM uses multiple-feature learning modules to learn multi-sequence embedding feature vectors. FGCNN complements CNN with MLP to learn global-local feature interactions, which has two components: feature generation and deep classifier. In the feature generation component, CNN is used to learn local feature interactions that are recombined by MLP to generate global feature interactions. In the deep classifier component, deep learning structures (e.g., PIN, xDeepFM) can be used to predict CTR. FGCNN can outperform CCPM and deep learning models (e.g., DeepFM, xDeepFM, IPNN) in AUC and Logloss. DMCNN introduces a dense matrix of quantum physics to address the shortcoming of CNN only capturing local neighboring patterns, which is composed of three layers: an embedding layer, a DMCNN layer and a deep layer. The embedding layer transforms the high-dimensional sparse feature space into a low-dimensional dense vector; the DMCNN layer translates the embedding vectors into a density matrix, then uses CNN to extract low-order feature interactions. The deep layer concatenates the output of the normalized embedding layer and the DMCNN layer to extract high-order feature interactions and obtains the prediction value. DMCNN outperforms CCPM for CTR prediction in AUC and Logloss, which implies that both high-order and low-order feature interactions are significant predictors.

In order to extract useful and important feature interactions from high-dimensional and sparse features, Lian & Ge (2020) proposed fine-grained feature interaction network (FINET) which leverages CNN to capture high-order feature interactions and the attention mechanism to learn feature importance in an interpretable manner. Chen et al. (2016a) used CNN to extract visual features automatically from image advertisements for CTR prediction. Edizel et al. (2017) leveraged cross-convolutional operators to obtain high-order feature representations of queries and advertisements in sponsored search advertising.

### 4.3.3 Factorization Machine supported Neural Network (FNN)

Deep neural networks (DNN) are effective for high-order feature representations (Wang et al., 2017; Ling et al., 2017). In general, a DNN has a feature input layer, more than two hidden layers processing input features, and an output layer delivering the predicted probability through an activation function. Naturally, a higher-order feature representation can be achieved by increasing the number of DNN layers.

Although DNN can represent high-order features and thus obtain better prediction performance (Ouyang et al., 2019a; Livne et al., 2020; Luo et al., 2020), it cannot be directly applied in a high-dimensional input feature space because of its expensive computation for training a large number of parameters. Zhang et al. (2016) proposed a factorization machine supported neural network (FNN) model for CTR prediction, where a pre-trained FM is taken to obtain dense vectors that serve as the input of DNN. It is worthwhile to note that Huang et al. (2017) explored an inverted modeling structure of FNN, which utilizes a DNN to learn high-order nonlinear features offline and then use FM to make online CTR prediction. The FNN modeling framework is shown in Figure 9.

In FNN, the FM is pre-trained to obtain parameters of linear terms $w_i$ $(i = 1,2, \dots, n)$ and of interaction terms $\boldsymbol{v_i}(i = 1,2, \dots, n)$ through the following formulation:

$$F_{FM}(\boldsymbol{w}, \boldsymbol{v}, \boldsymbol{x}) = w_0 + \sum_{i=1}^{n} w_i x_i + \sum_{i=1}^{n} \sum_{j=i+1}^{n} \langle \boldsymbol{v_i}, \boldsymbol{v_j} \rangle x_i x_j. \quad (9)$$

Then the input vector of DNN is constructed in the following way:

$$\begin{cases} \boldsymbol{z} = (w_0, \boldsymbol{z_1}, \boldsymbol{z_2}, \dots, \boldsymbol{z_n}) \\ \boldsymbol{z_i} = \boldsymbol{W_0^i} \boldsymbol{x_i}, i = 1,2, \dots, n \end{cases}, \quad (10)$$

where $w_0$ is the global bias, $\boldsymbol{W_0^i} \in \mathbb{R}^{(k+1) \times size(x_i)}$ is a parameter matrix which maps field features $\boldsymbol{x_i}$ to a $(k+1)$-dimensional vector. The first row of $\boldsymbol{W_0^i}$ is $w_i$, and the $(j+1)-$ $th$ row corresponds to the $j-th$ component of $\boldsymbol{v}_i$, i.e., $v_i^j$, $j = 1,2,\ldots,n-1$.

The input of DNN can be initialized as $\boldsymbol{z_i} = (w_i, v_i^1, v_i^2, \ldots, v_i^n), i = 1,2,\ldots,n$. For the purpose of illustration, consider a DNN with two hidden layers. The first hidden layer of DNN is fully connected with each field. Let $\boldsymbol{z} \in \mathbb{R}^{n\times(k+1)+1}$ denote the input of the first hidden layer. The best performing activation function for FNN is $tanh(\cdot)$. The first hidden layer is given as $\boldsymbol{L_1} = tanh(\boldsymbol{W_1 z} + \boldsymbol{b_1})$, where $\boldsymbol{W_1} \epsilon \mathbb{R}^{M\times[n\times(k+1)+1]}$, $\boldsymbol{b_1} \epsilon \mathbb{R}^M$, $M$ is the number of neurons in the first hidden layer. Similarly, the second hidden layer is also fully connected with the first hidden layer (Wang & He, 2018), which is given as $\boldsymbol{L_2} = tanh(\boldsymbol{W_2 L_1} + \boldsymbol{b_2})$, where $\boldsymbol{W_2} \epsilon \mathbb{R}^{L\times M}$, $\boldsymbol{b_2} \epsilon \mathbb{R}^L$, $L$ is the number of neurons in the second hidden layer. The click probability $p = \sigma(\boldsymbol{W_3 L_2} + b_3)$, where $\boldsymbol{W_3} \in \mathbb{R}^{1\times L}$, $\boldsymbol{b_3} \in \mathbb{R}$.

FNN can represent high-order feature interactions and learn valuable patterns from categorical feature interactions. Moreover, FNN reduces computational complexity for high-dimensional prediction problems. In addition, as illustrated by experiments on CTR prediction by Zhang et al. (2016), FNN performs better than LR and FMs in AUC. However, FNN has several shortcomings: (a) the efficiency of FNN is limited by the pre-trained FM (Guo et al., 2017); (b) embedding parameters may be excessively affected by the pre-trained FM; (c) FNN does not include low-order feature interactions which may be also important for CTR prediction (Zhang et al., 2021a).

Another valuable effort for strengthening FMs with high-order feature representations is product-based neural networks (PNN) (Qu et al., 2016; 2018). Specifically, PNN consists of three components, including an embedding layer for distributed representation of categorical features, a product layer for interactive representation between fields, and a fully connected DNN for high-order feature interactions. Structurally, PNN extends FNN by adding a product layer. PNN has three variants, namely IPNN, OPNN and PNN$*$, corresponding to three product operations (inner product, outer product, inner-outer product), respectively. However, since the product layer and the input layer of DNN are fully connected, computational complexity of PNN is high. Moreover, similar to FNN, PNN does not include low-order features.

### 4.3.4 DeepFM

In order to represent both low-order and high-order feature interactions in a unified framework, it is of necessity to combine DNN with low-order modeling frameworks discussed in Sections 4.1 and 4.2. A notable effort in this direction is DeepFM (Guo et al., 2017; 2018), which combines DNN and FM for CTR prediction. Specifically, in the DeepFM modeling framework, as illustrated in Figure 10, an embedding layer transforms sparse input features into dense features, which serve as the shared input of DNN and FM. In DeepFM, DNN and FM are combined to learn low-order features and high-order features simultaneously.

DeepFM has several advantages: (a) it can realize low-order feature interactions explicitly and high-order feature interactions implicitly in a unified framework; (b) it shares the raw feature input among FM and DNN components; (c) it does not require pre-training FMs and manual feature engineering, thus can achieve end-to-end learning; (d) the deep component of DeepFM can be replaced with other types of deep network architectures such as PNN (Guo et al., 2018). As reported by Guo et al. (2017), DeepFM performs better than LR, FMs, PNN (OPNN, IPNN, PNN*) and Wide & Deep for CTR prediction in AUC and LogLoss.

In this research branch, researchers have developed a few CTR prediction models similar to the modeling structure of DeepFM, e.g., Wide & Deep (Cheng et al., 2016), Wide & ResNet (Gao & Bie, 2018), attention stacked autoencoder (ASAE) (Wang et al., 2018), eXtreme Deep Factorization Machine (xDeepFM) (Lian et al., 2018), hybrid feature fusion (HFF) (Shi & Yang, 2020) and FO-FTRL-DCN (Huang et al., 2020). Wide & Deep is an interesting attempt for the sparse problem, which joins LR with a feed-forward neural network; however, it depends on business knowledge as LR requires manual feature engineering. Wide & ResNet adopts LR as the wide component and the residual network as the deep component to represent highly nonlinear features. ASAE utilizes an AFM component to obtain low-order feature interactions and a SAE (stacked autoencoder) component to capture high-order feature interactions; among the two components the input is shared. Because ASAE uses an improved FM for low-order feature interactions, it also can achieve feature interactions in an end-to-end manner, without feature engineering based on raw features. xDeepFM combines CIN and DNN, and the common input is shared among the two components. In xDeepFM, CIN achieves explicit bounded-degree feature interactions, rather than explicit second-order feature interactions by FM in DeepFM. xDeepFM outperforms DeepFM on CTR prediction datasets in AUC and Logloss (Lian et al., 2018). The HFF model consists of two layers, namely a feature interaction layer to capture high-order features and a feature fusion layer to fuse low- and high-order features with the multi-head self-attention mechanism. FO-FTRL-DCN combines Deep & Cross network (DCN) (Wang et al., 2017) and the optimization technique of follow the regularized leader (FTRL) (McMahan, 2011) to realize explicit and implicit high-order feature representations, and the prediction is conducted by a LR combination layer of the two components. FO-FTRL-DCN performs better than DCN for CTR prediction in AUC and Logloss (Huang et al., 2020).

In the literature (e.g., Jiang et al., 2016; 2018; Liu et al., 2017; She & Wang, 2018), various deep architecture models such as deep belief network (DBN), Bayesian neural network (BNN), convolutional neural network (CNN) and stacked denoising autoencoder (SDAE) have also been integrated with low-order models for CTR prediction.

## 4.4 Tree models

Tree models have a wide range of applications in computational advertising (He et al., 2014; Zhang et al., 2014a; Carreón et al., 2019). Following the idea of boosting in ensemble learning, Gradient boosting decision tree (GBDT) and XGBoost are developed on the basis of gradient boosting machine (GBM), which have shown considerable successes for CTR prediction (Trofimov et al., 2012; He et al., 2014; Wang et al., 2016; Chen & Guestrin, 2016).

### 4.4.1 Gradient Boosting Decision Tree (GBDT)

GBDT combines the additive model and the forward stage-wise algorithm to form a tree model of machine learning (Friedman, 2001; 2002; Wang et al., 2016). GBDT is a nonlinear model which can address regression and classification problems with appropriate loss functions (Richardson et al., 2007; Dave & Varma, 2010; Trofimov et al., 2012; Ren et al., 2020). Figure 11 presents the GBDT modeling framework.

In GBDT, the target is to find a function $F^*(\boldsymbol{x})$ which minimizes the loss function $L$ (Ren et al., 2020):

$$F^*(\boldsymbol{x}) = \underset{F}{argmin}\ \ E_{\boldsymbol{x},y} L\big(y, F(\boldsymbol{x})\big). \qquad (11)$$

In a prediction learning problem, $y$ and $F(\boldsymbol{x})$ denote the observed and predicted values of the input $\boldsymbol{x}$, respectively. Problem (11) can be transformed into an approximation problem where boosting approximates $F^*(\boldsymbol{x})$ through an additive expansion (Trofimov et al., 2012), given as follows.

$$F_m(\boldsymbol{x}) = F_{m-1}(\boldsymbol{x}) + \beta_m h(\boldsymbol{x}, \boldsymbol{c_m}), \qquad (12)$$

where $h(\boldsymbol{x}, \boldsymbol{c_m})$ is called base leaner (or weaker learner), representing a function of $\boldsymbol{x}$ with a parameter vector $\boldsymbol{c_m}$, which can be seen as a regression tree in GBDT; $\beta_m$ is the expression coefficient, i.e., the weight of $m-th$ base learner.

Given that an initial guess $F_0(\boldsymbol{x})$ is determined, the prediction problem in Equation (11) can be transformed into the following numerical optimization problem (Dembczynski et al., 2008).

$$(\boldsymbol{c_m}, \boldsymbol{\beta_m}) = \underset{c,\beta}{argmin}\ L\big(y, F_{m-1}(\boldsymbol{x}) + \beta_m h(\boldsymbol{x}, \boldsymbol{c_m})\big), \quad (13)$$

where $L\big(y, F_{m-1}(\boldsymbol{x}) + \beta_m h(\boldsymbol{x}, \boldsymbol{c_m})\big)$ is a function of $F_{m-1}(\boldsymbol{x})$; $\beta_m h(\boldsymbol{x}, \boldsymbol{c_m})$ denotes the best greedy step toward $F^*(\boldsymbol{x})$ and $h(\boldsymbol{x}, \boldsymbol{c_m})$ is the direction of the steepest descent step in SGD which can be appropriated by $-[\partial_{F(x)} L\big(y, F(\boldsymbol{x})\big)]_{F(x)=F_{m-1}(x)}$.

Through exploring the optimal solution $\beta_m^*$ and $\boldsymbol{c}_{\boldsymbol{m}}^*$, a strong learner can be obtained with base learners at each step, as illustrated in Figure 11. Equation (12) can be reformulated as $F_m(\boldsymbol{x}) = F_{m-1}(\boldsymbol{x}) + \beta_m^* h(\boldsymbol{x}, \boldsymbol{c}_{\boldsymbol{m}}^*)$.

GBDT has several advantages (Wang et al., 2016): (a) it can handle multiple types of features, collinearity, non-sparse data processing and automatic feature selection; (b) it is invulnerable to missing data and missing functions; (c) GBDT is more interpretable in that it explicitly characterizes contributions of different features. However, there are several disadvantages of GBDT (Natekin & Knoll, 2013): (a) it suffers from the high memory consumption because all the base learners in the ensemble must be evaluated; (b) it is slower due to its sequential learning manner; (c) it performs ineffectively on sparse categorical data.

In order to tackle massive advertisements on Facebook, He et al. (2014) proposed the concatenation of GBDT and LR for CTR prediction, where the output of the GBDT is used as the input of LR. Such that, GBDT significantly increases the performance of LR through automatic feature selection. However, it is worth mentioning that, after feature processing in GBDT, features are transformed into a vector containing 0 and 1, which might be highly sparse in a high-dimensional space. To address this problem, several ensemble models have been proposed, e.g., GBDT+gcForest (Qiu et al., 2018) where GBDT is used to extract useful features, then cascades with gcForest to make CTR prediction, and GBDT+DNN (Ke et al., 2019) integrating GBDT and DNN to handle sparse categorical features and dense numerical features, respectively. The two ensemble models can achieve end-to-end training.

### 4.4.2 XGBoost

XGBoost is an extension of GBDT. That is, GBDT takes the first gradient of loss function to update the base learner, while XGBoost takes both the first gradient and the second gradient to solve the sparse problem (Chen and Guestrin, 2016; Ren et al., 2020). As a result, compared to GBDT, XGBoost is sparsity-aware. Moreover, GBDT constructs a new tree through fitting a negative gradient, while XGBoost does it through finding a new objective function. In addition, XGBoost can prevent the overfitting problem by adding a regularization term to the loss function.

Similar to GBDT, the XGBoost modeling framework uses the additive function $f_m(\boldsymbol{x})$ to predict the output, where a series of functions need to be estimated in order to approximate the prediction value, by minimizing the regularized loss function (Chen & Guestrin, 2016) which is given as

$$\begin{cases} L(p) = \sum_{i=1}^{N} l(y_i, p_i) + \sum_{m=1}^{M} \Omega(f_m) \\ p(\boldsymbol{x}) = \sum_{m=1}^{M} f_m(\boldsymbol{x}) \\ \Omega(f_m) = rL_m + \frac{1}{2}\lambda \parallel \boldsymbol{w} \parallel^2 \end{cases}, \quad (14)$$

where $M$ is the number of base learners $f_m(\boldsymbol{x})$, $p_i(p(\boldsymbol{x_i}))$ is the predicted value of the $i-th$ instance by the $M-th$ tree model, $l(y_i, p_i)$ is the loss of instance $\boldsymbol{x_i}$, and $\Omega(f_m)$ penalizes the modeling complexity of tree models (Moneera et al., 2021), $L_m$ is the number of leaf nodes of the $m-th$ base learner, and $\boldsymbol{w}$ is the weight vector of leaf nodes.

Similarly to GBDT, $p(x)^{(m)}$ is used to solve the prediction problem in Equations (14) numerically. $p(x)$ can be reformulated as $p(x)^{(m)} = p(x)^{(m-1)} + f_m(x)$. Then the loss function of the $m-th$ tree model can be given as

$$L^m(p^{(m-1)}) = \sum_{i=1}^{N} l\left(y_i, p^{(m-1)}(\boldsymbol{x_i}) + f_m(\boldsymbol{x_i})\right) + \Omega(f_m). \quad (15)$$

Different from GBDT, XGBoost uses the second-order Taylor expansion at ( $y_i$ , $p(x)^{(m-1)}$) to approximate the above loss function (Wang et al., 2018), which is given as

$$L^m(p^{(m-1)}) \approx \sum_{i=1}^{N}[l\left(y_i, p^{(m-1)}(\boldsymbol{x_i})\right) + g_{m,i} f_m(\boldsymbol{x_i}) + \frac{1}{2} h_{m,i} f_m^2(\boldsymbol{x_i})] + \Omega(f_m), (16)$$

where $g_{m,i} = \frac{\partial l\left(y_i, p^{(m-1)}(x)\right)}{\partial p^{(m-1)}(x)}$, $h_{m,i} = \frac{\partial^2 l\left(y_i, p^{(m-1)}(x)\right)}{\partial^2 p^{(m-1)}(x)}$ are the first and second gradient of the loss function, respectively.

As the tree model is composed of leaf nodes and corresponding weights, $f_m(\boldsymbol{x})$ can be expressed as $f_m(\boldsymbol{x}) = \boldsymbol{w_{m,q(x)}}$ (Ren et al., 2020), where $q(\boldsymbol{x})$ represents the structure of the tree, which maps instances $\boldsymbol{x}$ to leaf nodes $l$ $\{l = 1, 2, \dots, L_m\}$. Define $s_{m,l} = \{i \mid q(\boldsymbol{x_i}) = l\}$ as the instance set of leaf node $l$ in the $m-th$ tree. The optimal weight $w_{m,l}^*$ minimizes the loss function $L^m$ in Equation (16), which is obtained as $w_{m,l}^* = -\frac{G_{m,l}}{H_{m,l}+\lambda}$ , where $G_{m,l} = \sum_{i \in s_{m,l}} g_{m,i}$ , and $H_{m,l} = \sum_{i \in s_{m,l}} h_{m,i}$. Based on $w_{m,l}^*$, the corresponding optimal $L^{m*} = -\frac{1}{2}\sum_{l=1}^{L_m} \frac{{G_{m,l}}^2}{H_{m,l}+\lambda} + \gamma L_m$.

In general, it is impossible to enumerate all trees structures. Chen & Guestrin (2016) used a greedy algorithm to fit the loss function $L^{m*}$ through a procedure starting from a single leaf node and iteratively adding branches. As illustrated in Table 5 at each step split, the loss reduction can be represented as $-\frac{1}{2}\frac{(G_L+G_R)^2}{H_L+H_R+\lambda}+\gamma L_m-[-\frac{1}{2}\frac{{G_L}^2}{H_L+\lambda}-\frac{1}{2}\frac{{G_R}^2}{H_R+\lambda}+\gamma(L_m+1)]$, where $G_L$, $G_R$ are the sum of the first gradient of instance sets of left and right leaf nodes, respectively, after the split. The loss reduction equation can be used for evaluating the split candidates. Notably, parallel computing in the splitting can speed the model exploration (Ren et al., 2020). Chen & Guestrin (2016) compared XGBoost with R.gbm (Ridgeway, 2007) and scikit-learn (Pedregosa et al., 2011), showed that (a) XGBoost has a comparable performance with scikit-learn in AUC, but with a shorter time to train per tree; (b) XGBoost and scikit-learn outperform R.gbm in AUC.

However, GBDT and XGBoost underperform in cases with a high dimension and large data size (Ke et al., 2017). To overcome this defect, researchers (e.g., Meng et al., 2016; Ke et al., 2017) developed LightGBM whose performance is improved significantly over GBDT and XGBoost with a faster computing speed and less memory consumption. Specifically, LightGBM employs gradient-based one-side sampling (GOSS) and exclusive feature bundling (EFB) to deal with large data instances and features, and the leaf-wise tree growth strategy and the maximum depth of the tree are used to ensure high efficiency while preventing the overfitting problem. Shi et al. (2019) proposed an embedded model with XGBoost and FwFM for CTR prediction, which can obtain better prediction accuracy with fewer parameters. An & Ren (2020) proposed XGBoost deep factorization machine-supported neutral network (XGBDeepFM), which utilizes XGBoost+LR for feature selection, FMs for second-order feature interactions, and DNN for high-order feature representations. XGBDeepFM performs better than state-of-art models (e.g., linear, linear+FM, linear+DNN, DeepFM) in AUC and efficiency.

# 5. Discussions

In this section we first discuss CTR prediction models with respect to the order of feature interactions, and then turn to performance evaluation on various datasets.

## 5.1 Summary of CTR Prediction Models

According to the order of feature interactions, the CTR prediction models discussed in Section 4, can be categorized into two groups: low-order ($\leq 2$) and high-order ($> 2$) models. Table 6 compares modeling frameworks, with respect to the complexity, the order of feature interactions, advantages and disadvantages.

The former category includes statistical methods (i.e., LR and Poly2) and FMs-based models (i.e., FMs, FFMs and FwFMs). Among statistical methods (Section 4.1), LR simply uses individual features, and Poly2 models take combinational features as additional inputs. FMs-based models (Section 4.2) are designed to make CTR prediction in the case with sparse data, among which FMs are equipped with second-order feature interactions, FFMs consider the difference of feature interactions between different pairs of fields, and FwFMs combine advantages of FMs and FFMs, reducing the number of parameters and the computational complexity while ensuring the prediction performance.

The latter category includes deep learning models (i.e., LSTM, CNN, FNN and DeepFM) and tree models (i.e., GBDT and XGBoost). Among deep learning-based models, LSTM is suitable for predicting problems with long-term dependencies, CNN favors local feature interactions and reduces the number of parameters, FNN and PNN cascade FMs and DNN to achieve high-order feature interactions, Wide & Deep and DeepFM realize explicit low-order feature interactions and implicit high-level feature interactions simultaneously. GBDT and XGBoost are two representative tree modeling frameworks based on boosting ideas, which can continuously reduce the loss by increasing base classifiers (i.e., new trees). GBDT uses the first-order gradient of the current model to minimize the loss function, while XGBoost uses the first and second-order gradient.

## 5.2 Datasets and Model Evaluation

In the extant literature, advertising CTR prediction models have been evaluated based on a bunch of benchmark datasets. Table A2 in the appendix presents publicly available datasets and proprietary datasets. We can notice that Criteo-Kaggle display advertising challenge 2014, Avazu and KDDCUP 2012 track 2 are the three most frequently used public datasets, and proprietary datasets are collected from advertising platforms in social media (e.g., Facebook) and e-commerce (e.g., Taobao). It is worthwhile to note that public datasets are more popular than proprietary datasets, possibly due to data accessibility and/or research replicability.

Generally, it is feasible to control modeling parameters to measure the performance of a class of prediction models on a same dataset. In prior studies (e.g., Qu et al., 2016; 2018; Guo et al., 2017; 2018), researchers have assessed impacts of neural network architecture parameters (e.g., embedding size, network depth, network width, network shape, activation function, dropout rate, learning rate) on the prediction performance across several models on a same dataset. These experimental comparisons primarily serve the following purposes: selecting parameters for the best prediction performance (Juan et al., 2016; Zhang et al., 2016;

Pan et al., 2018; Qu et al., 2016) and demonstrating the consistent superiority of a specific model (or a class of models) over others (Guo et al., 2017; 2018).

Performance evaluation of advertising CTR prediction models proposed in the literature on various datasets is summarized in Table 7. For both online advertising platforms and advertisers, it is critical to predict the advertising CTR accurately and quickly (Richardson et al., 2007; Graepel et al., 2010; Zhu et al., 2010; Agarwal et al., 2009; Hillard et al., 2010). From Table 7, we can obtain several observations.

First, we examine experimental comparisons between multivariate statistical models (i.e., LR and Poly2) and FMs reported in the literature on CTR prediction. Among 62 studies conducted on 7 public datasets and 2 proprietary datasets, most reported that FMs perform better than LR and Poly2 in AUC, Logloss, RMSE, accuracy, RelaImpr and RIG, with a few exceptions found that FMs underperform LR (Shi & Yang, 2020; Qu et al., 2016; Li et al., 2019; 2021b; Xie et al., 2021) and Poly2 (Chang et al., 2020).

Second, as for performance comparisons between FMs and their major extensions (i.e., FFMs and FwFMs), we can observe that almost all studies provided positive evidences for theoretical improvement of modeling extensions. Specifically, (a) among 28 studies conducted on 4 public datasets and 1 proprietary dataset, most reported that FFMs outperform FMs in AUC, Logloss, RMSE and accuracy, while Guo et al. (2019) and Khawar et al. (2020) provided an opposing result; (b) among 8 studies conducted on 2 public datasets and 1 proprietary dataset, 6 studies reported that FFMs outperform FwFMs in AUC and Logloss, while 2 studies found FwFMs perform better than FFMs when using the same number of parameters.

Third, through scrutinizing comparisons between deep models and FMs, we can notice strong proofs for the superiority of deep models. Specifically, (a) among 18 studies conducted on 5 public datasets, all except for Yan et al. (2020), reported that FNN outperforms FMs in in AUC, Logloss, RMSE and RIG; (b) among 26 studies conducted on 6 public datasets, all except for Lian et al. (2018), reported that PNN outperforms FMs in AUC, Logloss, RMSE, accuracy and RIG.

Fourth, comparisons between deep models showed that DeepFM and its extension (i.e., xDeepFM) outperform a couple of models (e.g., FNN, Wide & Deep, CCPM). Specifically, (a) among 17 studies conducted on 4 public datasets and 1 proprietary dataset, most reported that DeepFM performs better than FNN in AUC, Logloss, RelaImpr and RMSE, with a few exceptions (Huang et al., 2019; Lian & Ge, 2020; Niu & Hou, 2020; Zou et al., 2020b; Khawar et al., 2020); (b) among 17 studies conducted on 4 public datasets and 2 proprietary datasets, most reported that DeepFM performs better than Wide & Deep in AUC, Logloss, RelaImpr and RMSE, with a few exceptions (Ke et al., 2019; Ouyang et al. 2019a; Lian & Ge, 2020; Wu et al., 2020; Khawar et al., 2020); (c) All 7 studies conducted on 3 public datasets consistently reported that DeepFM performs better than CCPM in AUC and Logloss. In addition, the improvement of xDeepFM over DeepFM is largely confirmed: among 23 studies conducted on 6 public datasets, most reported that xDeepFM perform better than DeepFM in AUC and Logloss, with a few exceptions (Huang et al., 2019; Li et al., 2020b; Shi & Yang, 2020; Li et al., 2021a; Meng et al., 2021).

Moreover, we can observe that PNN is superior to FNN. Specifically, among 14 studies conducted on 4 public datasets and 1 proprietary dataset, most reported that PNN performs better than FNN in AUC, Logloss, RMSE and RIG, with a few exceptions (Guo et al., 2018; Yan et al., 2020; Huang et al., 2020). However, the literature provides controversial results on performance comparison between DeepFM and PNN (or its extensions). Specifically, among 28 studies conducted on 6 public datasets and 1 proprietary dataset, 16 studies reported that PNN performs better than DeepFM in AUC, Logloss, RMSE and accuracy, while 6 studies provided opposing results, and 6 studies gave inconsistent results. A closer inspection showed that, 13 studies explicitly used IPNN for comparisons, among which 11 studies found IPNN is superior and 2 studies supported DeepFM; however, the other 15 studies did not mention which variant of PNN was employed in experiments.

In summary, high-order models usually perform better than low-order models. Although the majority of studies provided theoretically consistent results on modeling performance, there were a few inconsistent and even opposing evidences. Especially, the most controversial results have been reported in regard to performance comparison between DeepFM and PNN. The possible explanations are given as follows. Data pre-processing and feature engineering (especially manual operations) significantly influence model performance, which may work quite differently across studies. Meanwhile, a model may not perform better in terms of all evaluation metrics, even on a same dataset. In other words, a model with a higher AUC value may not necessarily have a smaller Logloss than others (Liu et al., 2019; Huang et al., 2019). In addition, a model may favor some data characteristics that are different across datasets, thus perform differently on different datasets. As we can notice, several studies (e.g., Yan et al., 2020; Huang et al., 2019; Guo et al., 2018; Shi & Yang, 2020; Lian et al., 2018; Lian & Ge, 2020) repeatedly produced results against mainstream findings. We believe that comparison studies across various CTR prediction models certainly deserve further investigation.

# 6. Research Perspectives

In this section, we discuss current research trends, main challenges and potential future directions in the domain of advertising CTR prediction.

## 6.1 Current Research Trends

### 6.1.1 Powerful Feature Interactions in FMs

In order to capture potential correlations among various features, exploring powerful feature interactions has become a critical issue for CTR prediction. Indeed, interactive features promisingly enhance the prediction performance. In order to achieve explicit modeling and great interpretability, matrix weighting and attention mechanisms can be used to represent novel second-order interaction terms in the FMs framework (Chen et al., 2019; Wang et al., 2020c; Sun et al., 2021) and upgrade FMs with high-order and even arbitrary-order equipments (Tao et al., 2020; Yu et al., 2020).

### 6.1.2 Handling Numerical Features

As CTR prediction involves massive categorical features, the majority of modeling frameworks focuses on capturing interactions among categorical features. However, the embedding of numerical features has been largely overlooked. For the purpose of handling numerical features for CTR prediction, Guo et al. (2021a) proposed an embedding framework with meta-embeddings for each numerical field to obtain shared knowledge, differentiable discretization to capture correlations between embeddings and features, and an aggregation function to learn continuous representation for each feature.

Another possible solution is to transform numerical features into categorical values by employing tree models (He et al., 2014) and then search for embedding dimensions so as to obtain embeddings for these features (Zhao et al., 2020b; Liu et al., 2021).

### 6.1.3 Ensemble Modeling Frameworks

In ensemble modeling for CTR prediction, a notable research stream took GBDT as a major component for selecting useful features to feed learning components, e.g., GBDT+gcForest (Qiu et al., 2018), GBDT+DNN (Ke et al., 2019) and XGBDeepFM (An & Ren, 2020).

Another stream aims to design an ensemble form of low- and high-order components capable of capturing feature interactions of multiple orders flexibly (Zhu et al., 2020). In recent years, ensemble deep-learning models composed of FMs and DNNs have become a popular issue (e.g., He & Chua, 2017; Wang et al., 2018; Zou et al., 2020a; Yang et al., 2020).

### 6.1.4 Parallelism Schemes

Almost all deep CTR prediction models adopt an embedding layer to transform the high-dimensional sparse input into low-dimensional dense vectors, which easily results in memory requirements reaching hundreds of GB or even TB (Guo et al., 2021c). This demands parallelism schemes to speed up model implementation, which are widely adopted in deep learning systems (Jia et al., 2018, Chen et al., 2018).

In order to reduce communication costs in parallelization, Gupta et al. (2021) proposed a compression framework called dynamic communication thresholding (DCT) in large-scale hybrid training for CTR prediction, which compresses parameter gradients during model synchronization for data parallelism, and squeezes activations and gradients across sub-networks during the forward and backward propagations for model parallelism. In large-scale training systems, the latency of pull & push operations between CPU (Host) and GPU workers is a serious concern for media providers. Guo et al. (2021c) developed a MixCache-based distributed training system for CTR prediction, in which embedding tables are stored in Host memory of servers and GPU is utilized to conduct embedding parameter synchronization.

## 6.2 Main Challenges

### 6.2.1 Feature Engineering for CTR Prediction

Most of existing CTR prediction models rely heavily on manual feature engineering (e.g., LR, DNN). However, manual feature engineering not only prevents research replicability, but also is hard to implement in large-scale applications. It calls for automatic feature learning techniques, with little manual participation in implementations.

Although existing CTR prediction models have been ameliorated through feature interactions, they still suffer from several shortcomings in feature engineering. First, explicit feature interactions (e.g., FMs-based models) mainly concentrate on the low-order models resulting in better model interpretability, while implicit feature interactions (e.g., DNN-based models) favor high-order models but hamper the model interpretability. In practice, unexplainable CTR predictions may be unreliable. In a recent research, Li et al. (2020b) presented a transformer-based CTR prediction method with hierarchical attention layers for feature learning, providing interpretable insights yielded from prediction results. Second, it is apparent that prediction models with explicit high-order feature interactions (e.g., DCN and CIN models) demand more concrete theoretical supports. Third, not all combinatorial features can improve the model performance. In other words, it is noticeable that feature interactions unnecessarily make prediction models perform better (Ren et al., 2020). For models with automatic feature interactions, especially high-order feature interactions (e.g., PNN, DeepFM), it needs to conduct the feature importance analysis over combinatorial features by using attention and pooling methods (Zhang et al., 2021b). Additionally, existing models use the vector-product to represent interactions between each pair of features, while ignoring different semantic spaces (e.g., preference spaces of users and advertisers) (Wu et al., 2020).

### 6.2.2 Sample Imbalance between Clicks and Non-clicks

Generally, in the context of online advertising, the number of non-clicks is far more than the number of clicks due to the fact that fewer users give positive feedbacks to advertisements displayed to them. This raises the problem of imbalance distribution of sample labels (Feng et al., 2014; Liu et al., 2018; Zhang et al., 2017). That is, classifiers may perform well on the majority class (i.e., non-click samples) but poor on the minority class (i.e., click samples) (Zou et al., 2016).

In this direction, Xie et al. (2019) and Jiang et al. (2021) explored CTR prediction based on transfer learning to leverage common characteristics of click and non-click samples. However, the samples of advertising CTR are highly heterogeneous (Deng et al., 2017), which makes it unstraightforward to sufficiently span the complete sample space.

### 6.2.3 CTR Prediction for New Advertisements: Cold Start

Typically, there is little historical data for CTR prediction in the case with newly launched advertisements, i.e., the cold-start problem. That is, there is rare information for model training (Chakrabarti et al., 2008; Ouyang et al., 2021; Roy and Guntuku, 2016).

For deep learning models, it is impossible to obtain a good embedding vector for newly-added advertisements or those with a small size of training samples. To this end, Pan et al. (2019) proposed a meta-embedding model to address the cold start problem of new advertisements by exploiting attributes associated with them. However, it may leave out valuable information contained in other related advertisements, page layouts and users' behaviors (Ouyang et al., 2019a; 2019b; 2019c).

### 6.2.4 Data Sparsity and Heterogeneity for CTR Prediction

Advertising data usually contain a large number of categorical features which need binary representation, resulting in high dimensional vectors with many zero elements (Gharibshah & Zhu, 2021). This raises the data sparsity problem for CTR prediction. However, most of CTR prediction models, such as Poly2, DNN, cannot be directly applied on sparse data (Chang et al., 2020; Pan et al., 2018; Zhang et al., 2016). Due to the sparse input, embedding parameters occupy a large computational space, which declines the convergence rate of prediction models (Yan et al., 2020; Shi et al., 2020).

Moreover, advertising contents and users' behaviors are highly heterogeneous so that the related information might be unobserved in limited samples (Deng et al., 2017; Ying, 2019). Advertising heterogeneity raises the overfitting problem for CTR prediction. That is, prediction models perform well on the training dataset, but less well on the testing dataset. In supervised machine learning, overfitting cannot be completely avoided because of the limits of training data (e.g., the limited size and high noises) and algorithms (e.g., the complexity and the large set of parameters) (Ying, 2019). Moreover, advertising heterogeneity encumbers the generalization of CTR prediction models.

## 6.3 Future Directions

### 6.3.1 Graph Neural Networks (GNN)

Owning to its advantages such as expressive power for graph data, high interpretation and ground-breaking performance, Graph neural network (GNN) becomes an effective framework for graphical representation, analysis and learning. In recent years, several research attempts have been reported employing GNN for CTR prediction (Li et al., 2019; 2021b), focusing on simple forms of feature interactions using graphic representation.

We believe that GNN can be used to integrate more powerful feature interactions (e.g., FwFM, FmFM, AOAFM) in the graph structure and apply diverse aggregate strategies to achieve better performance for CTR prediction. Moreover, GNN has the potential to address the cold-start problem by building a graph connecting various advertisements, and distilling useful information from neighboring advertisements to boost CTR prediction for new advertisements (Ouyang et al., 2021).

### 6.3.2 Neural Architecture Search

In most modeling frameworks for CTR prediction, a unified embedding dimension is empirically allocated to all feature fields, which is memory inefficient and limits the prediction performance. Neural architecture search (NAS) enables adaptively and automatically searching for an embedding dimension of each field according to its contribution to prediction (Cheng et al., 2020; Zhao et al., 2020b; Liu et al., 2021).

In recent years, there emerged a tendency of efficiently searching for feature interactions in network architectures. In other words, it aims to automatically identify useful feature interactions from all possible feature combinations and select an appropriate structure to capture these interactive features (Luo et al., 2019; Liu et al., 2020; Zhao et al., 2020a; Khawar et al., 2020). Moreover, NAS can be viewed as a good technology in automatically constructing modeling frameworks (Wang et al., 2020b; Meng et al., 2021; Liu et al., 2021). It is suggested to create search space with diversified blocks, design flexible interaction rules, and develop efficient search strategies in neural architectures (Song et al., 2020).

### 6.3.3 Explicit Representation of High-order Feature Interactions

Researchers have invested substantial efforts into exploring explicit high-order models for CTR prediction, among which Deep & Cross Network (DCN) (Wang et al., 2017), compressed interactions network (CIN) (Lian et al., 2018) and interaction machine (IM) (Yu et al., 2020) are three meaningful attempts in this direction. Unfortunately, explicit representation and high interpretation of these models are achieved at the cost of prediction performance (Lian et al., 2018; Huang et al., 2019; Yan et al., 2020). Another remarkable way is using graph neural network (GNN) to intuitively represent feature interactions in the graph structure (Li et al., 2019; 2021b; Xie et al., 2021; Guo et al., 2021b).

In the direction of explicit high-order models, there are two promising perspectives. The first perspective is to replace the DNN component with explicit high-order models (e.g., DCN, CIN, IM) in deep-learning modeling frameworks (e.g., DeepFM) for CTR prediction. The second perspective is to model field-aware feature interactions in explicit high-order modeling frameworks, following principles of FFMs and FwFMs. That is, it is suggested to consider impacts of multi-field interactions on the prediction performance when increasing the order of feature interactions.

### 6.3.4 Understanding User Behaviors

In online advertising, understanding user behaviors can significantly facilitate preference elicitation and CTR prediction. This calls for smart user models to guide sample preprocessing and feature engineering. For example, directly treating non-clicks as negative samples may lead to a serious noise issue, because users who are interested in advertisements may not click. In a recent work, Zhao et al. (2021) proposed a noise filtering approach based on reinforcement learning to identify effective negative samples.

Another interesting issue in this direction is to capture users' interests by mining the sequential behavior information. In this direction, Zhou and his colleagues explored a set of attention-based CTR models to learn temporal user interests from rich behavior sequences toward specific advertisements (Zhou et al., 2019; Pi et al., 2020). Considering the drift of user interests over time, it calls for more elegant models tackling the temporal correlation and spillover effects in user behaviors and social networking interactions among users in social media advertising.

### 6.3.5 Advertising Characteristics

On the Internet, a large number of novel advertising forms distinguish from each other with inherent characteristics. Among these characteristics, advertising position is significantly associated with CTR in sponsored search advertising (Yang et al., 2018). In this sense, the position information (Yuan et al., 2020; Huang et al., 2021) and the externality effect from surrounding advertisements (Deng et al., 2018) are definitely useful to improve prediction results. In this direction, it calls for more attention on the power of advertising characteristics in CTR prediction.

### 6.3.6 CTR Prediction for Multimedia Ads

Multimedia advertising has become popular in recent years, possibly due to its strengthen in attracting consumers' attentions and the rapid development of Internet. Existing CTR prediction models only consider categorical features and numerical features, while ignoring image and video features that are distinct characteristics in multimedia advertising.

Previous research on multimodal learning focuses on extracting and fusing multimodal features through concatenation (Cheng et al., 2012; Zhou et al., 2019; Chen et al., 2020) and dynamically assigning weights for each modality through attention mechanisms (Li et al., 2018). It is promising to explore interactions between different modalities in multimedia advertising, which is an important component in CTR prediction tasks.

## 7. Conclusion

Online advertising has been becoming a major way of delivering marketing and advertising messages to potential consumers since the last decade. Advertising CTR prediction will remain as a hot research topic in the coming decades. In this paper, we present a comprehensive literature review on advertising CTR predication, with special focuses on modeling frameworks. Moreover, we discussed advantages and disadvantages of state-of-the-art CTR prediction models and performance evaluation. Lastly, current research trends, main challenges and potential future directions worthy of further exploration in this area are summarized.

In addition to future directions discussed in Section 6.3, it is worthwhile to conduct comparison studies evaluating various CTR prediction models on benchmark datasets, in order to reconcile inconsistent results reported in the literature (Section 5.2). In the literature, there are about 175 CTR prediction models developed and/or experimentally compared over 9 public datasets and dozens of proprietary datasets in the articles included in this review. Another challenge for comparison studies lies in the design of computational protocols including data pre-processing, sampling, feature engineering and experimental setups.

This review is expected to provide fundamental knowledge and efficient entry points for IS and marketing scholars who want to engage in this area. In the meanwhile, this review may present the theoretical basis for modeling frameworks for CTR prediction, facilitating development and implementation of novel models.

**Table 1. Notations**

| Terms | Definition |
|---|---|
| $y_i$ | Target label: $y_i = 1$ means click, $y_i = 0$ means don't click. |
| $p_i$ | $p_i \in [0,1]$ is the predicted click probability of instance $i$. |
| $N$ | Sample size. |
| $L$ | Loss function. |
| $l$ | Single sample loss |
| $\nabla l(w)$ | The gradient of single sample loss. |
| $L_m$ | The number of the leaf nodes of $m - th$ tree. |
| σ | Activation function. |
| w | Weight. |
| $b$ | Bias term. |
| $n$ | The number of features. |
| $m$ | The number of feature fields. |
| $H$ | the dimension of hashing space in Poly2 |
| $k$ | the dimension of embedding vectors |
| $q(x)$ | The function which maps the instance to leaf node of tree. |
| $i_t$ | The input gate of LSTM cell. |
| $f_t$ | The forget gate of LSTM cell. |
| $o_t$ | The output gate of LSTM cell. |
| $h_t$ | The hidden state of LSTM cell. |
| TP | True positive |
| FP | False positive |
| TN | True negative |
| FN | False negative |

**Table 2. Feature Types for CTR prediction**

| Type | Feature examples |
|---|---|
| Advertising | ad ID, advertiser ID, campaign ID, creative ID, ad keyword, title, body, URL domain, type of ad, ad display position, advertiser network, conversion id, ad group, ad size, ad length, ad width, etc. |
| User | gender, age, region, user ID, user agent, IP, accept cookies, etc. |
| Context | city, carrier, time, network speed, device brand, device type, OS, OS name, OS version, etc. |
| Query | query category, query keywords, query term, average query length, etc. |
| Publisher | publisher ID, publisher network, site, section, URL, page referrer, etc. |

**Table 3. Evaluation Metrics for CTR Prediction**

| Evaluation Metrics | Definition | References |
|---|---|---|
| Precision | $P = \frac{TP}{TP+FP}$ | Chakrabarti et al., (2008); Wang and Chen (2011); Wang et al. (2016); Qiu et al. (2018); Gharibshah et al. (2020) |
| Recall | $R = \frac{TP}{TP+FN}$ | Chakrabarti et al., (2008); Wang and Chen (2011); Wang et al. (2016); Qiu et al. (2018); Gharibshah et al. (2020) |
| F1-score | $F_1 = \frac{2\times TP}{N+TP-TN}$ | Wang and Chen (2011); Wang et al. (2016); Qiu et al. (2018); Gharibshah et al. (2020) |
| Accuracy | $\text{accuracy} = \frac{TP+TN}{N}$ | Wang and Chen (2011); Phangtriastu & Isa (2018); Zhang et al. (2019); Xie et al. (2019); Gharibshah et al. (2020); Jiang et al. (2021); Gupta et al. (2021); |
| AUC-ROC | ROC curve is characterized by the ratio of True Positive Rate (TPR) and False Positive Rate (FPR), i.e., TPR/ FPR. $TPR = \frac{TP}{TP+FN}$ $FPR = \frac{FP}{TN+FP}$ AUC-ROC is the area under the receiver operating characteristics (ROC) curve. | Cheng et al. (2012); Rendle (2012b); Wang et al. (2013); McMahan et al. (2013); Zhang et al. (2014a); Zhang et al. (2014b); Ta (2015); Li et al. (2015); Meng et al. (2016); Chen & Guestrin (2016); Shan et al. (2016); Pan et al. (2016); Chen et al. (2016a); Zhang et al. (2016); Qu et al. (2016); Wang et al. (2016); Jiang et al. (2016); Guo et al. (2017); Jie-Hao et al. (2017); Ke et al. (2017); Ling et al. (2017); Zhang et al. (2017); Huang et al. (2017); Liu et al. (2017); jiang et al. (2017); Edizel et al. (2017); Pan et al. (2018); Qiu et al. (2018); Wang et al. (2018); Wang & He (2018); Zhou et al. (2018); Guo et al. (2018); Chan et al. (2018); Liu et al. (2018); Deng et al. (2018); Lian et al. (2018); Jiang et al. (2018); Gao & Bie (2018); She & Wang (2018); Pi et al. (2019); Shi et al. (2019); Ke et al. (2019);Ouyang et al. (2019a); Ouyang et al. (2019b); Ouyang et al. (2019c); Liu et al. (2019); Li et al. (2019); Yuan et al. (2019); Gligorijevic et al. (2019); Zhang et al. (2019); Huang et al. (2019); Wu et al. (2019); Qu et al. (2019); Guo et al. (2019); Wang et al. (2019); Xie et al. (2019); Zhou et al. (2019); Luo et al. (2019); Chen et al. (2019); Pan et al. (2019); Zhang et al. (2020); Gharibshah et al. (2020); Luo et al. (2020); An & Ren (2020); Chang et al. (2020); Chen et al. (2020); Tao et al. (2020); Huang et al. (2020); Wang et al. (2020a); Niu & Hou (2020); Zou et al. (2020b); Lian & Ge (2020); Shi & Yang (2020); Li et al. (2020a); Li et al. (2020b); Liu et al. (2020); Livne et al. (2020); Yan et al. (2020); Song et al. (2020); Yang et al. (2020); Pan et al. (2020); Pi et al. (2020); Yu et al. (2020); Yuan et al. (2020); Wu et al. (2020); Zhao et al. (2020a); Zhao et al. (2020b); Zhu et al. (2020); Khawar et al. (2020); Li et al. (2021a); Jiang et al. (2021); Sun et al. (2021); Ouyang et al. (2021); Guo et al. (2021a); Guo et al. (2021b); Huang et al. (2021); Li et al. (2021b); Liu et al. (2021); Meng et al. (2021); Xie et al. (2021); Zhao et al. (2021) |
| AUC-PR | PR curve is characterized by the ratio of precision and | Chapelle et al. (2014) |
| | recall, i.e., precision / recall. AUC-PR is the area under the curve of precision/recall. | |
| Relative improvement (RelaImpr) | $RelaImpr = \left[\frac{AUC(model)-0.5}{AUC(baseline)-0.5} - 1\right] \times 100\%$ | Zhou et al. (2018); Chan et al. (2018); Chen et al. (2020); Zhao et al. (2021) |
| Logloss | $L = -\frac{1}{N}\sum_{i=1}^{N}[\ y_i \times logp_i + (1 - y_i)(1 - logp_i)]$ | Trofimov et al. (2012); Zhang et al. (2014b); Liu et al. (2015); Chen et al. (2016a); Qu et al. (2016); Pan et al. (2016); Wang et al. (2016); Chen et al. (2016b); Juan et al. (2016); Guo et al. (2017); Wang et al. (2017); Liu et al. (2017); Jie-Hao et al. (2017); jiang et al. (2017); Juan et al. (2017); Gao & Bie (2018); Guo et al. (2018); Liu et al. (2018); Lian et al. (2018); Wang et al. (2018); She & Wang (2018); Zhou et al. (2018); Ouyang et al. (2019a); Ouyang et al. (2019b); Ouyang et al. (2019c); Liu et al. (2019); Zhang et al. (2019); Qu et al. (2019); Gligorijevic et al. (2019); Huang et al. (2019); Guo et al. (2019); Wang et al. (2019); Zhou et al. (2019); Chen et al. (2019); Pan et al. (2019); Li et al. (2019); Zhang et al. (2020);Chang et al. (2020); Li et al. (2020a); Li et al. (2020b); Liu et al. (2020); Tao et al. (2020); Wang et al. (2020a); Huang et al. (2020); Niu & Hou (2020); Livne et al. (2020); Zou et al. (2020b); Lian & Ge (2020); Shi & Yang (2020); Yan et al. (2020); Song et al. (2020); Yang et al. (2020); Zhao et al. (2020b); Pan et al. (2020); Yu et al. (2020); Shi et al. (2020); Wang et al. (2020b); Zhao et al. (2020a); Zhu et al. (2020); Li et al. (2021a); Sun et al. (2021); Ouyang et al. (2021); Guo et al. (2021a); Guo et al. (2021b); Gupta et al. (2021); Li et al. (2021b); Liu et al. (2021); Meng et al. (2021); Guo et al. (2021c) |
| Mean square error (MSE) | $MSE = \frac{1}{N}\sum_{i=1}^{N}(p_i - y_i)^2$ | Richardson et al. (2007); Dembczynski et al. (2008); Xiong et al. (2012); Trofimov et al. (2012); Qu et al. (2019); Wang et al. (2020b) |
| Root Means Squared Error (RMSE) | $RMSE = \sqrt{\frac{\sum_{i=1}^{n}(y_i - p_i)^2}{N}}$ | Zhang et al. (2014a); Avila Clemenshia & Vijaya (2016); Pan et al. (2016); Qu et al. (2016); Shan et al. (2016); Wu et al. (2019); Wang et al. (2020a); Li et al. (2020a); Wang et al. (2020c); Yang et al. (2020); Cheng et al. (2020) |
| Relative Information Gain (RIG) | $RIG = 1 - \frac{LL_{predict}}{LL_{empirical}}$ $LL_{predict} = -\frac{1}{N}\sum_{i=1}^{N}[y_i logp_i + (1 - y_i)\log(1 - p_i)]$ $LL_{empirical} = -\frac{1}{N}\sum_{i=1}^{N}[y_i logp_e + (1 - y_i)\log(1 - p_e)]$ | He et al. (2014); Zhang et al. (2014b); Qu et al. (2016); Ling et al. (2017); Deng et al. (2018) |
| Field-level Calibration Error | Field-level expected calibration error (Field-ECE):<br>$Field - ECE = \frac{1}{\lvert\mathcal{D}\rvert}\sum_{z=1}^{\lvert\mathcal{Z}\rvert}\left\lvert\sum_{i=1}^{\lvert\mathcal{D}\rvert}(y_i - \hat{p}_i)1_{[z_i=z]}\right\rvert$<br><br>Field-level relative calibration error (Field-RCE):<br>$Field - RCE = \frac{1}{\lvert\mathcal{D}\rvert}\sum_{z=1}^{\lvert\mathcal{Z}\rvert} N_z \frac{\left\lvert\sum_{i=1}^{\lvert\mathcal{D}\rvert}(y_i - \hat{p}_i)1_{[z_i=z]}\right\rvert}{\sum_{i=1}^{\lvert\mathcal{D}\rvert}(y_i + \varepsilon)1_{[z_i=z]}}$ | Pan et al. (2020) |

**Table 4. Categories of CTR Prediction Models in the Literature.**

| Category | Modeling Approach | References |
|---|---|---|
| Multivariate statistical models | LR | Richardson et al. (2007); Chakrabarti et al., (2008); Xiong et al., (2012); Wang et al. (2013); Yan et al. (2014); Zhang et al. (2014a); Zhang et al. (2014b); Chapelle et al. (2014); Liu et al. (2015); Kumar et al. (2015); Avila Clemenshia & Vijaya (2016); Zhang et al. (2016); Ma et al. (2016); Jiang et al. (2016); Chen et al. (2016a); Chen et al. (2016b); Wang et al. (2016); Shan et al. (2016); Qu et al. (2016); Jie-Hao et al. (2017); Jiang et al. (2017); Wang et al. (2017); Ling et al. (2017); Zhang et al. (2017); Edizel et al. (2017); Guo et al. (2018); Gao & Bie (2018); Jiang et al. (2018); Zhou et al. (2018); Liu et al. (2018); Shi et al. (2019); Liu et al. (2019); Wu et al. (2019); Guo et al. (2019); Yuan et al. (2019); Ouyang et al. (2019a); Ouyang et al. (2019b); Ouyang et al. (2019c); Qu et al. (2019); Ke et al. (2019); Wang et al. (2019); Huang et al. (2020); Niu & Hou (2020); Zhang et al. (2020a); Yu et al. (2020); Lian & Ge (2020); Yan et al. (2020) |
| | Poly2 | Ma et al. (2016); Jie-Hao et al. (2017); Pan et al. (2018); Guo et al. (2018); Chang et al. (2020) |
| Factorization machines (FMs) based models | FM | Rendle (2012b); Ta (2015); Liu et al. (2015); Qu et al. (2016); Zhang et al. (2016); Pan et al. (2016); Ma et al. (2016); Juan et al. (2017); Jie-Hao et al. (2017); Huang et al. (2017); Wang et al. (2017); Liu et al. (2017); Pan et al. (2018); Guo et al. (2018); Jiang et al. (2018); Wang et al. (2018); Wang & He (2018); Liu et al. (2018); Phangtriastu & Isa (2018); She & Wang (2018); Shi et al. (2019); Ouyang et al. (2019a); Ouyang et al. (2019b); Ouyang et al. (2019c); Ke et al. (2019); Liu et al. (2019); Huang et al. (2019); Zhang et al. (2019); Pan et al. (2019); Li et al. (2019); Yuan et al. (2019); Guo et al. (2019); Chen et al. (2019); Wang et al. (2019); Huang et al. (2020); Liu et al. (2020); Wang et al. (2020c); An & Ren (2020); Chang et al. (2020); Tao et al. (2020); Niu & Hou (2020); Zou et al. (2020b); Lian & Ge (2020); Li et al. (2020b); Yu et al. (2020); Yan et al. (2020); Yang et al. (2020); Sun et al. (2021); Li et al. (2021b); Liu et al. (2021); Xie et al. (2021) |
| | FFM | Ma et al. (2016); Jie-Hao et al. (2017); Pan et al. (2018); Wang & He (2018); Liu et al. (2018); Qu et al. (2018); Phangtriastu & Isa (2018); Zhang et al. (2019); Shi et al. (2019); Liu et al. (2019); Yuan et al. (2019); Guo et al. (2019); Chen et al. (2019); Huang et al. (2020); Luo et al. (2020); Wang et al. (2020c); Chang et al. (2020); Yan et al. (2020); Yang et al. (2020); Sun et al. (2021) |
| | FwFM | Pan et al. (2018); Shi et al. (2019); Chen et al. (2019); Sun et al. (2021) |
| Deep learning models | LSTM | Chen et al. (2016b); Deng et al. (2018); Qu et al. (2019); Gligorijevic et al. (2019); Zhou et al. (2019); Pi et al. (2019); Zhang et al. (2020); Gharibshah et al. (2020); Wang et al. (2020a) |
| | CNN | Liu et al. (2015); Chen et al. (2016a); Edizel et al. (2017); Chan et al. (2018); Liu et al. (2019); Lian and Ge (2020); Niu & Hou (2020) |
| | FNN | Zhang et al. (2016); Qu et al. (2016); Guo et al. (2018); Wang et al. (2018); Wang & He (2018); Huang et al. (2020); Niu & Hou (2020); Zou et al. (2020b); Lian & Ge (2020); Yan et al. (2020) |
| | DeepFM | Guo et al. (2017); Guo et al. (2018); Jiang et al. (2018); Lian et al. (2018); Zhou et al. (2018); Liu et al. (2018); Zhang et al. (2019); Ouyang et al. (2019a); Liu et al. (2019); Ouyang et al. (2019c); Ke et al. (2019); Pan et al. (2019); Guo et al. (2019); Wang et al. (2019); Wang et al. (2020a); Chen et al. (2019); Zhang et al. (2020); Niu & Hou (2020); Zou et al. (2020b); Li et al. (2020a); Lian & Ge (2020); Li et al. (2020b); Liu et al. (2020); Livne et al. (2020); Yan et al. (2020); Yu et al. (2020); Yang et al. (2020); Song et al. (2020); Li et al. (2021a); Ouyang et al. (2021); Liu et al. (2021) |
| Tree models | GBDT | Trofimov et al. (2012); He et al. (2014); Meng et al. (2016); Wang et al. (2016); Ke et al. (2017); Ling et al. (2017); Qiu et al. (2018); Ke et al. (2019); Liu et al. (2019) |
| | XGBoost | Chen & Guestrin (2016); Ke et al. (2017); Livne et al. (2020); An & Ren (2020) |

**Table 5. The splitting rules.**

| Splitting rule | $x \leq c$ | | $x > c$ | | |
|---|---|---|---|---|---|
| Instances | $x_1$ | $x_4$ | $x_2$ | $x_3$ | $x_5$ |
| The first gradient of instance | $g_1$ | $g_4$ | $g_2$ | $g_3$ | $g_5$ |
| The second gradient of instance | $h_1$ | $h_4$ | $h_2$ | $h_3$ | $h_5$ |
| $G_L$ , $G_R$ | $G_L = g_1 + g_4$ | | $G_R = g_2 + g_3 + g_5$ | | |
| $H_L$ , $H_R$ | $H_L = h_1 + h_4$ | | $H_R = h_2 + h_3 + h_5$ | | |

**Table 6. Comparison of CTR Prediction Models.**

| Models | Number of parameters | Order of feature interactions | Advantages | Disadvantages |
|---|---|---|---|---|
| LR | $n$ | 1 | (a) LR can obtain a value between 0 and 1 describing click probability;<br>(b) LR is a linear combination of various features, capturing correlations between features and the label;<br>(c) LR's gradient of loss function has an elegant form. | (a) LR fails to represent interactive effects and non-linear relationships among features and the label because it assumes each feature to be independent. |
| Poly2 | $n+H$ | 2 | (a) Poly2 can explicitly describe interactions between features.<br>(b) Compared to linear models, Poly2 performs better in CTR prediction in terms of accuracy. | (a) Compared to linear models, Poly2 is slower because of its higher complexity.<br>(b) Poly2 performs less well on sparse data. |
| FMs | $n+nk$ | 2 | (a) FMs perform better than LR and Poly2 where the data is sparse.<br>(b) FMs have a linear complexity in both the dimensionality of the factorization and the number of features. | (a) FMs fail to capture the fact that, when interacting with various features from multiple fields (e.g., gender and age), a feature might behave differently. |
| FFMs | $n+n(m-1)k$ | 2 | (a) FFMs are field-aware for feature interactions. | (a) FFMs have a large number of parameters that need to be estimated, which significantly increases the modeling complexity. |
| FwFMs | $n+nk+\frac{m(m-1)}{2}$ | 2 | (a) FwFMs can outperform LR, Poly2 and FMs, and achieve comparable performance with FFMs, while with fewer parameters to be estimated. | (a) FwFMs do not contain high order feature interactions. |
| LSTM | - | >2 | (a) LSTM can overcome the shortcoming of RNN when the data has long-term dependencies;<br>(b) LSTM can effectively solve the gradient vanishing and exploding problem when using SGD to find the optimal solution. | (a) LSTM needs more time and larger memory in the training step. |
| CNN | - | >2 | (a) CNN is powerful to find local feature interactions and reduce the number of parameters. | (a) The sequence of embedding feature vectors significantly affects local feature interactions in CNN;<br>(b) Useful feature interactions may be lost in CNN. |
| FNN | - | >2 | (a) FNN can represent high-order features and learn effective patterns from categorical feature interactions;<br>(b) FNN can reduce computational complexity of high-dimensional input feature problems. | (a) The efficiency of FNN is limited by the pre-trained FM;<br>(b) Embedding parameters may be excessively affected by the pre-trained FM;<br>(c) FNN does not include low-order feature representations. |
| DeepFM | - | >2 | (a) DeepFM can realize low-level feature interactions explicitly and high-order feature interactions implicitly in a unified framework;<br>(b) DeepFM shares the raw feature input among FM and DNN components;<br>(c) DeepFM does not require pre-training FMs and manual feature engineering, thus can achieve end-to-end learning;<br>(d) The deep component of DeepFM can be replaced with other types of deep network architectures such as PNN. | (a) The deep component of DeepFM makes high-order feature interactions less interpretable;<br>(b) DeepFM has a high computational complexity. |
| GBDT | - | >2 | (a) GBDT can handle multiple types of features, collinearity, non-sparse data processing and automatic feature selection;<br>(b) GBDT is invulnerable to missing data and missing functions;<br>(c) GBDT models are more interpretable in that they explicitly characterize contributions of different features. | (a) GBDT suffers from the high memory consumption because all the base learners in the ensemble must be evaluated;<br>(b) GBDT is slower to learn due to its sequential learning manner;<br>(c) GBDT underperforms in cases with high feature dimensions and big data;<br>(d) GBDT performs less well on sparse data. |
| XGBoost | - | >2 | (a) Compared to GBDT, XGBoost is sparsity-aware. | (a) XGBoost underperforms in cases with high feature dimensions and big data. |

**Table 7. Comparison for Model Performance on Different Datasets**

| Dataset | References | Model Performance |
|---|---|---|
| Criteo-Kaggle display advertising challenge 2014 | Wang et al. (2016) | AUC:<br>LR < RF < RT+LR < GBDT < RF+LR < GBDT+LR;<br>Logloss:<br>GBDT+LR < RF+LR < RT+LR < GBDT < RF < LR;<br>Precision:<br>LR < RF, RT+LR < GBDT < RF+LR < GBDT+LR;<br>Recall:<br>LR < RF < RT+LR < GBDT < RF+LR < GBDT+LR;<br>F1-score:<br>LR < RF < RT+LR < GBDT < RF+LR < GBDT+LR; |
| | Juan et al. (2016) | Logloss:<br>FFM < FM < Poly2-SG < LM-SG; |
| | Chen & Guestrin (2016) | AUC:<br>R.gbm < scikit-learn < XGBoost; |
| | Guo et al. (2017) | AUC:<br>LR < FM < FNN < IPNN < OPNN < PNN* < DeepFM;<br>Logloss:<br>DeepFM < PNN* < OPNN < IPNN < FNN < FM < LR; |
| | Wang et al. (2017) | Logloss:<br>DCN < DNN < FM < LR; |
| | Juan et al. (2017) | Logloss:<br>FFM < LR; |
| | Pan et al. (2018) | AUC:<br>LR < Poly2 < FMs < FwFM < FFM; |
| | Guo et al. (2018) | AUC:<br>LR < FM < DNN < IPNN < OPNN < PNN∗ < FNN < DeepFM;<br>Logloss:<br>DeepFM < FNN < PNN∗ < OPNN < IPNN < DNN < FM < LR; |
| | Qiu et al. (2018) | Precision:<br>Random Forest < SVM < Naïve Bayes < GBDT < ETCF;<br>Recall:<br>Naïve Bayes < Random Forest, SVM < ETCF < GBDT;<br>F1-score:<br>Random Forest, SVM < Naïve Bayes < GBDT < ETCF;<br>AUC:<br>Random Forest < Naïve Bayes < SVM < GBDT < ETCF; |
| | Phangtriastu & Isa (2018) | Accuracy:<br>FFM < PSO-FFM; |
| | Lian et al. (2018) | AUC:<br>LR < FM < DNN < Wide&Deep < DeepFM < DCN < PNN < xDeepFM;<br>Logloss:<br>xDeepFM < DCN < DeepFM < Wide&Deep < DNN < FM < LR < PNN; |
| | Gao & Bie (2018) | AUC:<br>LR < DNN < Wide&Deep < DenseNet < ResNet < Wide&ResNet; |
| | Ke et al. (2019) | AUC:<br>LR < FM < DeepFM < PNN < Wide&Deep < GBDT < DeepGBM; |

| | |
|---|---|
| Xie et al. (2019) | AUC:<br>SVM < Centroid < SMOTE < SGAN < DNN < ADASYN < DAN < PILKE < RTILKE; |
| Li et al. (2019) | AUC:<br>LR < FM < Deep&Cross < AFM < NFM < CIN < DeepCrossing < FiGNN;<br>Logloss:<br>Fi-GNN < DeepCrossing < CIN < NFM < AFM < Deep&Cross < LR < FM; |
| Chang et al. (2020) | AUC:<br>FM < Poly2 < FFM;<br>Logloss:<br>FFM < Poly2 < FM; |
| Tao et al. (2020) | AUC:<br>CIN < GBDT+LR < FM < Wide&Deep < xDeepFM < HoAFM;<br>Logloss:<br>HoAFM < xDeepFM < Wide&Deep < FM < GBDT+LR < CIN; |
| Yan et al. (2020) | AUC:<br>LR < FM < FFM < PNN < AFM < Wide&Deep < FNN < NFM < DCN < AutoInt < DeepFM < MoFM;<br>Logloss:<br>MoFM < AutoInt < DeepFM < DCN < NFM < Wide&Deep < PNN < FNN < AFM < FFM < FM < LR; |
| Li et al. (2020a) | Logloss:<br>ADIN < FNFM < DeepFM < AFM < DeepCross < FNN < Wide&Deep;<br>AUC:<br>Wide&Deep < FNN < DeepCross < AFM < DeepFM < FNFM < ADIN;<br>RMSE:<br>ADIN < FNFM < DeepFM < AFM < DeepCross < FNN < Wide&Deep; |
| Li et al. (2020b) | AUC:<br>FM < InterHAt-s < DCN < InterHAt, Wide&Deep < xDeepFM < DeepFM < PNN;<br>Logloss:<br>PNN < DeepFM < xDeepFM < Wide&Deep, InterHAt < DCN < InterHAt-S < FM; |
| Luo et al. (2020) | AUC:<br>FFM < NFM < AutoInt < Wide&Deep < DNN < xDeepFM < NON; |
| Wang et al. (2020c) | RMSE:<br>AOAFFM < FFM < AFM < FM; |
| Lian & Ge (2020) | AUC:<br>LR < NFM < Wide&Deep < FNN < DeepFM < PNN < XDeepFM < FINET;<br>Logloss:<br>FINET < XDeepFM < PNN < DeepFM, FNN < Wide&Deep < NFM < LR < FM; |
| Yang et al. (2020) | Logloss:<br>ONN < FFM < DNN < PNN < DeepFM <FM;<br>AUC:<br>FM < DeepFM <DNN < PNN < FFM < ONN;<br>RMSE:<br>ONN < FFM < DNN < DeepFM < PNN < FM; |

| | | |
|---|---|---|
| | Yu et al. (2020) | AUC:<br>LR < FM < HOFM < IM;<br>Logloss:<br>IM < HOFM < FM < LR; |
| | Shi & Yang (2020) | AUC:<br>LR < FM < AFM < NFM < Wide&Deep < AutoInt < DeepFM < Deep&Cross < xDeepFM < HFF;<br>Logloss:<br>HFF < xDeepFM, Deep&Cross < DeepFM < AutoInt < Wide&Deep < NFM < AFM < LR < FM; |
| | Song et al. (2020) | AUC:<br>DLRM < DeepFM < AutoInt+ < AutoCTR;<br>Logloss:<br>AutoCTR < AUtoInt+ < DLRM < DeepFM; |
| | Sun et al. (2021) | AUC:<br>LR < Wide&Deep < DeepFM < xDeepFM < AutoInt < FM < FwFM < FvFM < FFM, FiBiNET < FmFM < Deep&Cross < DeepLight;<br>Logloss:<br>DeepLight < FmFM < Deep&Cross < FFM < FvFM < xDeepFM < FiBiNET < FwFM < FM < AutoInt < DeepFM < LR < Wide&Deep; |
| | Li et al. (2021b) | AUC:<br>LR < FM < Deep&Cross <AFM < NFM < HOFM < DeepCrossing, xDeepFM < InterHAt <AutoINt <Fi-GNN < GraphFM;<br>Logloss:<br>GraphFM < Fi-GNN < AutoInt < InterHAt < HOFM < DeepCrossing < xDeepFM < NFM < AFM < Deep&Cross < LR < FM; |
| | Xie et al. (2021) | AUC:<br>Wide&Deep < DNN < FM < LR < AutoFIS < Fi-GNN < AutoInt < FIVES; |
| | Meng et al. (2021) | AUC:<br>LR < CrossNet < FM < AFM < HOFM < NFM < DeepFM < PNN < CIN < Deep&Cross < AFN < AutoInt < Wide&Deep < xDeepFM < AutoInt+ < AFN+ < AutoRec-R;<br>Logloss:<br>AutoRec-R < AFN+ < AutoInt+ < xDeepFM < Wide&Deep < AutoInt < AFN < Deep&Cross < CIN < DeepFM < PNN < NFM < HOFM < AFM < FM < CrossNet < LR; |
| Criteo | Qu et al. (2016) | AUC:<br>LR < FM < FNN < CCPM < PNN∗ < OPNN < IPNN;<br>Logloss:<br>IPNN < OPNN < CCPM < PNN∗ < FNN < LR < FM;<br>RMSE:<br>IPNN < OPNN < PNN∗< CCPM < FNN < FM < LR;<br>RIG:<br>LR < FM < FNN < PNN∗ < CCPM < OPNN < IPNN; |
| | Liu et al. (2018) | AUC:<br>LR < FM < FFM < DeepFM;<br>Logloss:<br>DeepFM < FFM < FM < LR; |
| | Huang et al. (2019) | AUC:<br>LR < FM < AFM < FFM < SE-FM-ALL;<br>AUC:<br>DCN < FNN < DeepFM < xDeepFM < DeepSE-FM-ALL; |
| | | Logloss:<br>SE-FM-ALL < FFM < AFM < FM < LR;<br>Logloss:<br>DeepSE-FM-ALL < DeepFM < xDeepFM < FNN < DCN; |
| | Liu et al. (2019) | AUC:<br>LR < GBDT < FM < FFM < CCPM < DeepFM < xDeepFM < IPNN < PIN < FGCNN;<br>Logloss:<br>FGCNN < PIN < IPNN < xDeepFM < DeepFM < FFM < CCPM < FM < GBDT < LR; |
| | Guo et al. (2019) | AUC:<br>LR < FM < DeepFM < FFM < xDeepFM < IPNN < PIN < OENN;<br>Logloss:<br>OENN < PIN < IPNN < xDeepFM < FFM < DeepFM < FM < LR; |
| | Niu & Hou (2020) | AUC:<br>LR < FM < FNN < CCPM < DeepFM < DMCNN;<br>Logloss:<br>DMCNN < FNN < DeepFM < CCPM < FM < LR; |
| | Zou et al. (2020b) | AUC:<br>LR < FM < Wide&Deep < DeepFM < FNN < PNN < INN;<br>Logloss:<br>INN < PNN < DeepFM < FNN < Wide&Deep < FM < LR; |
| | Liu et al. (2020a) | AUC:<br>FM < AFM < FwFM < GBDT+LR < AutoFM (2nd) < FM < GBDT+FFM < DeepFM < AutoDeepFM (2nd);<br>Logloss:<br>AutoDeepFM (2nd) < DeepFM < GBDT+FFM < FFM < AutoFM (2nd) < FwFM < FM < AFM < GBDT+LR; |
| | Wu et al. (2020) | AUC:<br>FM < AFM < NFM < DeepFM < Wide&Deep <TFNet; |
| | Khawar et al. (2020) | AUC:<br>LR < GBDT < FM < AFM < CCPM < FFM < FNN < DeepFM < xDeepFM < IPNN < PIN < AutoFeature(2) < AutoFeature(3) < AutoFeature(4);<br>Logloss:<br>AutoFeature(4) < AutoFeature(3) < PIN < AutoFeature(2) < IPNN < xDeepFM <DeepFM < FNN < FFM < CCPM < FM < AFM < GBDT < LR; |
| Avazu | Liu et al. (2015) | Logloss:<br>CCPM < FM < LR; |
| | Avila Clemenshia &Vijaya (2016) | RMES:<br>LR < SVR< PR; |
| | Chen et al. (2016b) | Logloss:<br>LSTM < RNN < LR < NN; |
| | Juan et al. (2016) | Logloss:<br>FFM < Poly2-SG < FM < LM-SG; |

| | |
|---|---|
| Liu et al. (2017) | AUC:<br>FTRL < ADPREDICTION < MATCHBOX < SPRASE-MLP, FM-MLP < FFM-MLP;<br>Logloss:<br>FM-MLP < FFM-MLP < SPRASE-MLP < MATCHBOX < ADPREDICTION < FTRL; |
| Chan et al. (2018) | AUC:<br>MSS-RD < MSS-SG <MSM-RD < MSM-SG;<br>RI:<br>MSS-RD < MSS-SG < MSM-RD < MSM-SG; |
| Liu et al. (2018) | AUC:<br>LR < FM < FFM < DeepFM;<br>Logloss:<br>DeepFM < FFM < FM < LR; |
| Qiu et al. (2018) | Precision:<br>Random Forest < SVM < Naïve Bayes < GBDT < ETCF;<br>Recall:<br>Random Forest < SVM < Naïve Bayes < GBDT < ETCF;<br>F1-score:<br>Random Forest < SVM < Naïve Bayes < GBDT < ETCF;<br>AUC:<br>SVM < Naïve Bayes < GBDT < Random Forest < ETFC; |
| She & Wang (2018) | AUC:<br>LR < FM < CNN < CNN+FM;<br>Logloss:<br>CNN+FM < CNN < FM < LR; |
| Huang et al. (2019) | AUC:<br>LR < AFM < FM < FFM < SE-FM-ALL;<br>Logloss:<br>SE-FM-ALL < FFM < FM < AFM < LR;<br>AUC:<br>DCN < DeepFM < FNN < XDeepFM < DeepSE-FM-ALL;<br>Logloss:<br>DeepSE-FM-ALL < FNN < DeepFM < XDeepFM < DCN; |
| Liu et al. (2019) | AUC:<br>LR < GBDT < FM < CCPM < FFM < DeepFM < xDeepFM < IPNN < PIN < FGCNN;<br>Logloss:<br>FGCNN < PIN < IPNN < xDeepFM < DeepFM < FFM < CCPM < FM < GBDT < LR; |
| Li et al. (2019) | AUC:<br>LR < DeepCrossing < FM < NFM < AFM < CIN < Deep&Cross < Fi-GNN;<br>Logloss:<br>Fi-GNN < CIN < AFM < FM < NFM < Deep&Cross < DeepCrossing < LR; |
| Shi et al. (2019) | AUC:<br>FM < FFM < XG-FFM < FwFM < XG-FwFM; |
| Zhang et al. (2019) | Logloss:<br>FNFM < DCN < FFM < DeepFM < PNN < NFM < FM < LR;<br>Accuracy:<br>LR < FM < FFM, PNN < DCN, DeepFM < NFM < FNFM;<br>AUC:<br>LR < FM < NFM < PNN < DeepFM < FFM < DCN < FNM; |
| Xie et al. (2019) | AUC:<br>SVM < Centroid < SMOTE < ADASYN < SGAN, DNN < DAN < PILKE < RTILKE; |
| Chang et al. (2020) | Logloss:<br>FFM < FM < Poly2;<br>AUC:<br>Poly2 < FM < FFM; |
| Chen et al. (2019) | AUC:<br>FFM < FwFM < DCN < DeepFM < NFM < xDeepFM < NFFM < FLEN;<br>Logloss:<br>FLEN < NFFM < xDeepFM < NFM < DCN < DeepFM < FwFM < FFM; |
| Guo et al. (2019) | AUC:<br>LR < FM < FFM < DeepFM < xDeepFM < IPNN < PIN < OENN;<br>Logloss:<br>OENN < PIN < IPNN < xDeepFM < DeepFM < FFM < FM < LR; |
| Lian & Ge (2020) | AUC:<br>LR < FM < DeepFM < FNN < Wide&Deep < PNN < NFM < XDeepFM < FINET;<br>Logloss:<br>FINET < NFM, PNN < XDeepFM < FNN < Wide&Deep < DeepFM < FM < LR; |
| Niu & Hou (2020) | AUC:<br>LR < FM < CCPM < DeepFM < FNN < DMCNN;<br>Logloss:<br>DMCNN < DeepFM < FNN < CCPM < FM < LR; |
| Tao et al. (2020) | AUC:<br>CIN < xDeepFM < Wide&Deep < GBDT+LR < FM < HoAFM;<br>Logloss:<br>HoAFM < FM < GBDT+LR < Wide&Deep < xDeepFM < CIN; |
| Yan et al. (2020) | AUC:<br>LR < FNN < AFM < FM < FFM < Wide&Deep < NFM < DeepFM < PNN < AutoInt < DCN < MoFM;<br>Logloss:<br>MoFM < DeepFM < AutoInt < Wide&Deep < DCN < PNN < NFM < FFM < FM < AFM < FNN < LR; |
| Yu et al. (2020) | AUC:<br>LR < FM < HOFM < IM;<br>Logloss:<br>HOFM < IM < FM < LR; |
| Li et al. (2020a) | Logloss:<br>ADIN < FNFM < DeepFM < AFM < DeepCross < FNN < Wide&Deep;<br>AUC:<br>Wide&Deep < FNN < DeepCross < AFM < DeepFM < FNFM < ADIN;<br>RMSE:<br>ADIN < FNFM < DeepFM < AFM < DeepCross < FNN < Wide&Deep; |
| Li et al. (2020b) | AUC:<br>FM < DeepFM < Wide&Deep, DCN < PNN, xDeepFM < InterHAt-S < InterHAt;<br>Logloss: |
| | InterHAt < PNN < xDeepFM < InterHAt-S < Wide&Deep, DeepFM < DCN < FM; |
| Liu et al. (2020) | AUC:<br>GBDT+LR < FM < AFM < FwFM < FFM, AutoFM (2nd) < GBDT+FFM < DeepFM < AutoDeepFM (2nd);<br>Logloss:<br>AutoDeepFM (2nd) < DeepFM < GBDT+FFM < AutoFM (2nd) < FFM < FwFM < AFM < FM < GBDT+LR; |
| Luo et al. (2020) | AUC:<br>DNN < Wide&Deep < AutoInt < NFM < xDeepFM < FFM < NON; |
| Moneera et al. (2021) | AUC:<br>LR < KNN < Random Forest < XGBoost; |
| Zou et al. (2020b) | AUC:<br>LR < FM < FNN < DeepFM < Wide&Deep < PNN < INN;<br>Logloss:<br>INN < DeepFM, PNN < Wide&Deep < FNN < FM < LR; |
| Wu et al. (2020) | AUC:<br>FM < AFM < DeepFM < NFM < Wide&Deep <TFNet; |
| Shi & Yang (2020) | AUC:<br>LR < FM < NFM < AFM < Deep&Cross < Wide&Deep < DeepFM < AutoInt < HFF < xDeepFM;<br>Logloss:<br>HFF < xDeepFM, AutoInt < Wide&Deep < DeepFM < Deep&Cross < AFM < FM < NFM< LR; |
| Song et al. (2020) | AUC:<br>DLRM < DeepFM <AutoInt+ < AutoCTR;<br>Logloss:<br>AutoCTR < AUtoInt+ < DLRM < DeepFM; |
| Khawar et al. (2020) | AUC:<br>LR < GBDT < FM < AFM < CCPM < FNN < FFM < DeepFM < xDeepFM < IPNN < PIN < AutoFeature(2) < AutoFeature(3) < AutoFeature(4);<br>Logloss:<br>AutoFeature(4) < AutoFeature(3) < AutoFeature(2) < PIN < IPNN < xDeepFM <DeepFM < FNN < FFM < AFM < CCPM < FM < GBDT < LR; |
| Sun et al. (2021) | AUC:<br>LR < FM < FwFM < FvFM <AutoInt < FFM < Fi-GNN < FmFM < Deep&Cross < FGCNN+IPNN < DeepLight;<br>Logloss:<br>FGCNN+IPNN < DeepLight < Deep&Cross < FmFM < AutoInt < Fi-GNN < FFM < FvFM < FwFM < FM < LR; |
| Jiang et al. (2021) | AUC:<br>LR < SVM < DSL < FM < MFT, CNN; |
| Li et al. (2021b) | AUC:<br>LR < DeepCrossing < Deep&Cross < HOFM < FM < NFM < AFM < InterHAt < AutoInt < xDeepFM < Fi-GNN < GraphFM;<br>Logloss:<br>GraphFM < AutoInt < Fi-GNN < xDeepFM < HOFM, AFM < NFM < FM < Deep&Cross < InterHAt < DeepCrossing < LR; |

| | | |
|---|---|---|
| | Meng et al. (2021) | AUC:<br>LR < AFM < AutoInt < AutoInt+ < FM < CrossNet < AutoRec-R < AFN < HOFM < PNN < Wide&Deep < NFM < CIN < xDeepFM < DeepFM < Deep&Cross < AFN+;<br>Logloss:<br>AFN+ < Deep&Cross < AFN < PNN < xDeepFM < FM < DeepFM < Wide&Deep < AutoRec-R < CrossNet < HOFM < CIN < NFM < AFM < AutoInt+ < AutoInt < LR; |
| | Zhao et al., (2021) | AUC:<br>LR < DIEN < DeepFM < YoutubeNet < Wide&Deep < FM < DIN;<br>RelaImpr:<br>DIEN < DeepFM < DIN < YoutubeNet < FM < Wide&Deep < LR; |
| Avito | Jiang et al. (2018) | AUC:<br>LR < FM < SNN < DeepFM < FMSDAELR; |
| | Ouyang et al. (2019a) | AUC:<br>LR < FM < DNN < Wide&Deep < DeepFM < GRU;<br>Logloss:<br>GRU < Wide&Deep < DeepFM < DNN < LR < FM; |
| | Ouyang et al. (2019b) | AUC:<br>LR < FM < DNN < Wide&Deep;<br>Logloss:<br>Wide&Deep < DNN < FM < LR; |
| | Ouyang et al. (2019c) | AUC:<br>LR < FM < DNN < PNN, Wide&Deep < DeepFM;<br>Logloss:<br>Wide&Deep < DeepFM < PNN < DNN < LR < FM; |
| | Xie et al. (2019) | AUC:<br>DNN < Centroid < ADASYN < SVM < DAN < SGAN < SMOTE < PILKE < RTILKE; |
| | Jiang et al. (2021) | AUC:<br>LR, FM < SVM < DSL < CNN <MFT; |
| KDD CUP 2012 track 2 dataset | Wang et al. (2013) | AUC:<br>LR < MCLPR < SVR; |
| | Jiang et al. (2017) | AUC:<br>LR < SVR < SAE-LR; |
| | Jie-Hao et al. (2017) | AUC:<br>LR < GBDT <FFM+FTRL; |
| | Ke et al. (2017) | AUC:<br>XGBoost (pre-sorted algorithm) < LightGBM; |
| | Wang et al. (2018) | AUC(Datasize:1000000):<br>FM < FNN < CCPM < Deep&cross < Wide&Deep < ASAE;<br>Logloss (Datasize:1000000):<br>ASAE < Wide&Deep < Deep&cross < CCPM < FNN < FM; |
| | Wang & He (2018) | AUC:<br>LR< GBDT < FNN; |
| | Wang et al. (2019) | AUC (Data size:1000000):<br>LR < FM < Wide&Deep < PNN < SAEFL;<br>DeepCross < DBNLR < NFM < DeepFM < SAEFL;<br>Logloss (Data size:1000000):<br>SAEFL < PNN < Wide&Deep < FM < LR;<br>SAEFL < DeepFM < NFM < DBNLR < DeepCross; |
| | Shi & Yang (2020) | AUC:<br>NFM < Wide&Deep < LR < AFM < FM < xDeepFM < DeepFM < Deep&Cross < AutoInt < HFF;<br>Logloss:<br>HFF < AutoInt < DeepFM, Deep&Cross < xDeepFM < FM < AFM < Wide&Deep < NFM< LR; |
| | Song et al. (2020) | AUC:<br>DeepFM < AutoInt+ < DLRM < AutoCTR;<br>Logloss:<br>AutoCTR < AUtoInt+, DLRM < DeepFM; |
| iPinYou | Zhang et al. (2014a) | AUC:<br>LR < GBRT;<br>Logloss:<br>GBRT < LR； |
| | Zhang et al. (2016) | AUC:<br>FM < LR < FNN; |
| | Shan et al. (2016) | AUC:<br>LR < FCTF;<br>RMSE:<br>FCTF < LR; |
| | Qu et al. (2016) | AUC:<br>LR < FM < FNN < CCPM < PNN∗ < IPNN < OPNN;<br>Logloss:<br>PNN∗ < IPNN < OPNN < FNN < FM < CCPM < LR;<br>RMSE:<br>PNN∗ < IPNN < FNN < OPNN < CCPM, FM < LR;<br>RIG:<br>LR < CCPM < FM < FNN < OPNN < IPNN < PNN∗; |
| | Pan et al. (2016) | RMSE:<br>SFM < BFM < FM;<br>NLL:<br>SFM < BFM < FM;<br>AUC:<br>BFM < FM < SFM; |
| | Guo et al. (2019) | AUC:<br>FFM < LR < FM < DeepFM < xDeepFM < IPNN < PIN < OENN;<br>Logloss:<br>OENN < PIN < IPNN < xDeepFM < DeepFM < FM < LR < FFM; |
| | Khawar et al. (2020) | AUC:<br>FFM < LR < GBDT < FM < CCPM < AFM < FNN < DeepFM < xDeepFM < IPNN < PIN < AutoFeature(2) < AutoFeature(3) < AutoFeature(4);<br>Logloss:<br>AutoFeature(4) < AutoFeature(3) < AutoFeature(2) < PIN < IPNN < xDeepFM < AFM < FNN < GBDT < DeepFM < FM < CCPM < LR < FFM; |
| | Huang et al. (2020) | AUC:<br>LR < FM < FFM < PNN < FNN < DCN;<br>Logloss:<br>DCN < FNN < PNN < FFM < FM < LR; |
| Aliyun Taobao display advertising dataset | Li et al. (2021a) | Logloss:<br>SESS-CAN-T < DeepFM < DCN < xDeepFM < MLP;<br>AUC:<br>MLP < xDeepFM < DeepFM < DCN < SESS-CAN-T; |
| | Guo et al. (2021b) | AUC:<br>LR < FM < FiGNN < AutoInt+ < GIN < DIN < DeepFM < DIEN <PNN < DG-ENN;<br>Logloss:<br>DG-ENN < PNN < DIEN < DeepFM < DIN < GIN <AutoInt+ < FiGNN < FM < LR; |
| Huawei DIGX algorithm contest dataset 2019 | Zhang et al. (2020) | AUC:<br>LR < DeepFM < ADPN;<br>Logloss:<br>ADPN < DeepFM < LR; |
| Taobao advertising dataset | Ouyang et al. (2021) | AUC：<br>DNN < Wide&Deep < AutoInt < DeepFM < PNN;<br>Logloss:<br>Wide&Deep < AutoInt < DeepFM < PNN < DNN; |
| Proprietary datasets | Trofimov et al. (2012) | Logloss:<br>MatrisNet < GBM < LR; |
| | He et al. (2014) | NE:<br>LR+Trees < LR only < Trees only; |
| | Zhang et al. (2014b) | AUC:<br>LR < NN < RNN;<br>RIG:<br>LR < NN < RNN; |
| | Yan et al. (2014) | RelaImpr:<br>LR < CGL; |
| | Ta (2015) | AUC:<br>FM < FTRFL; |
| | Chen et al. (2016a) | AUC:<br>LR < DNN < DeepCTR;<br>Logloss:<br>DeepCTR < DNN < LR; |
| | Jiang et al. (2016) | AUC:<br>SAELR < DBNLR; |
| | Zhang et al. (2017) | AUC:<br>LR < SVM < ELM; |
| | Huang et al. (2017) | AUC：<br>Dataset 1:LR < FM < DNN < DNN-LR <DSL < onlineDSL;<br>Dataset 2:LR < DNN < FM < DNN-LR <DSL < onlineDSL;<br>Dataset 3:LR < FM < DNN-LR < DNN <DSL < onlineDSL; |
| | Edizel et al. (2017) | AUC:<br>FELR < Search2Vec < DeepWordMatch < DeepCharMatch; |
| | Ling et al. (2017) | AUC Gain:<br>LR < LR+GBDT < NN < LR2GBDT < GBDT2NN < GBDT2LR < LR+GBDT < NN2GBDT < GBDT+DNN, NN+GBDT; |
| | Chan et al. (2018) | AUC:<br>MSS-RD < MSS-SG <MSM-RD < MSM-SG;<br>RI:<br>MSS-RD < MSS-SG < MSM-RD < MSM-SG; |
| | Pan et al. (2018) | AUC:<br>LR < Poly2 < FMs < FwFM < FFM; |
| | Zhou et al. (2018) | AUC:<br>LR < Wide&Deep < PNN < DeepFM < DIN;<br>RelaImpr:<br>LR < Wide&Deep < PNN < DeepFM < DIN; |
| | Qu et al. (2019) | AUC:<br>LR < Feedforward Neural Network < Wide&Deep < PNN < Dynamic Neural Network;<br>MSE：<br>Dynamic Neural Network < PNN < Wide&Deep < LR < Feedforward Neural Network;<br>Logloss:<br>Dynamic Neural Network < PNN < LR < Wide&Deep < Feedforward Neural Network; |
| | Ouyang et al. (2019b) | AUC:<br>LR < FM < DNN < Wide&Deep;<br>Logloss:<br>Wide&Deep < DNN < FM < LR; |
| | Zhou et al. (2019) | AUC:<br>Wide&Deep < PN < DIN < Two layers GRU Attention < DIEN; |
| | Gligorijevic et al. (2019) | AUC:<br>LM < VDCNN <MT <DM < DSM;<br>Accuracy:<br>MT < LM < DM < VDCNN < DSM; |
| | Pi et al. (2019) | AUC:<br>DIEN < MIMN; |
| | Gharibshah et al. (2020) | Precision:<br>Naïve Bayes < Random forest < DeepFM < LSTM < LR < CNN < Linear SVM < SVM;<br>Recall:<br>SVM < Linear SVM < Naïve Bayes < Random forest < LR <DeepFM < CNN < LSTM;<br>F1-measure:<br>SVM < Naïve Bayes < Random forest < Linear AVM < LR < DeepFM < CNN < LSTM;<br>AUC:<br>SVM < Linear SVM < Naïve Bayes < Random forest < LR < DeepFM < CNN < LSTM;<br>Accuracy:<br>LSTM < DeepFM < CNN < Naïve Bayes < Random forest < LR < Linear SVM < SVM; |
| | Ren et al. (2020) | Logloss:<br>XGBoost with FE (feature engineering) < LR with FE < DNN without FE < XGBoost without FE < LR without FE; |
| | An & Ren (2020) | AUC:<br>Wide&Deep < PNN, FNN < XDeepFM < XGBDeepFM. |
| | Zhao et al., (2021) | AUC:<br>LR < FM < DIN < DeepFM < DIEN < YoutubeNet < Wide&Deep;<br>RelaImpr:<br>Wide&Deep < YoutubeNet < FM < DIEN < LR < DIN < DeepFM; |

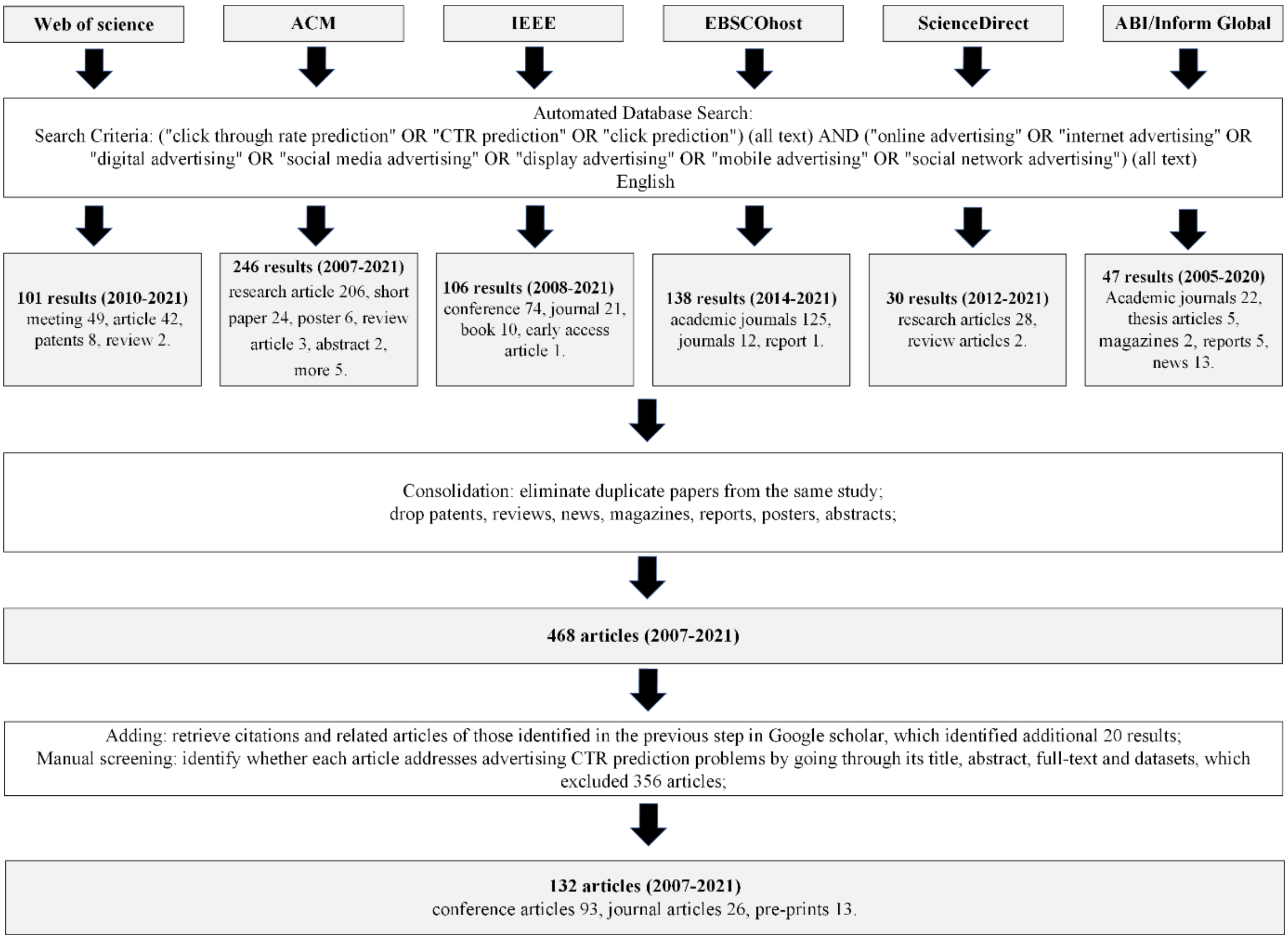

Figure 1. Study search and selection.

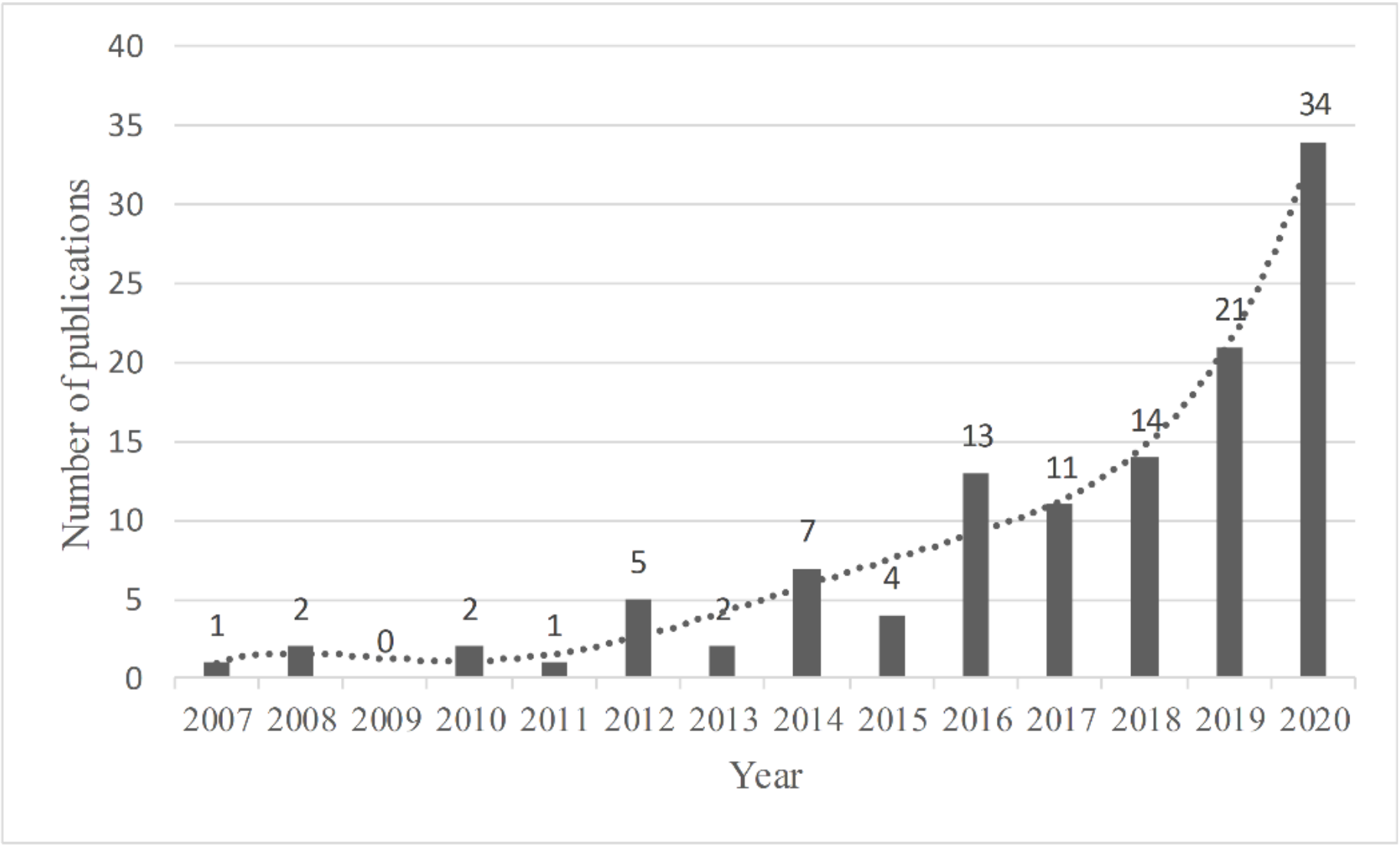

Figure 2. The distribution of selected studies on publication year.

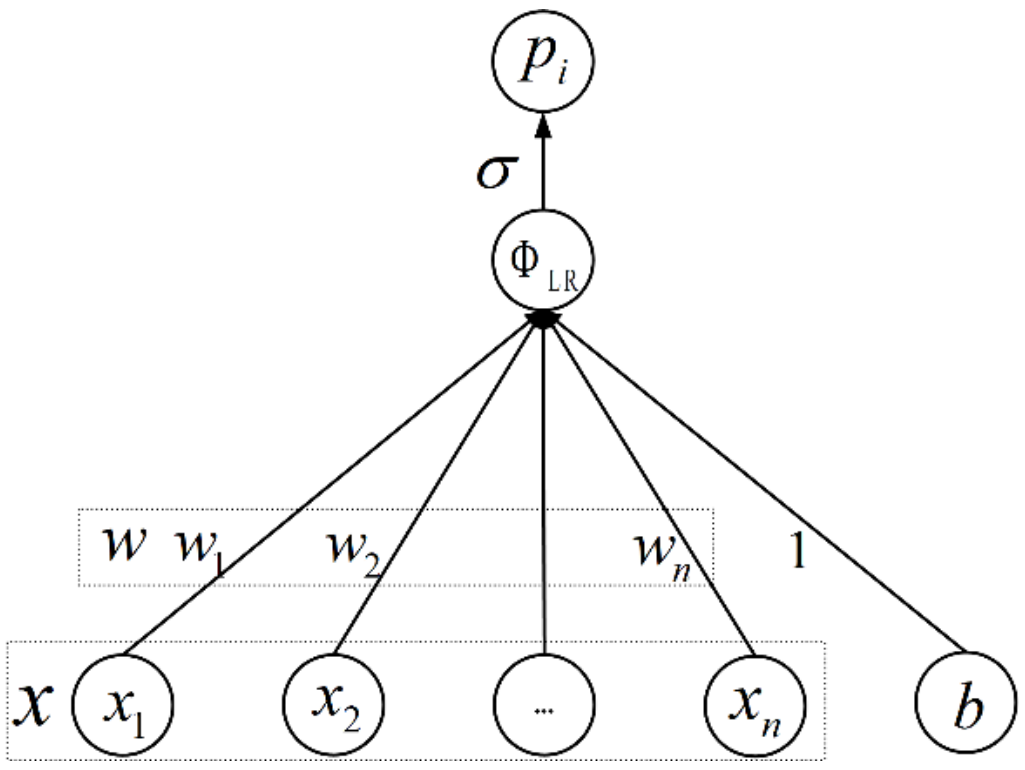

Figure 3. The LR modeling framework.

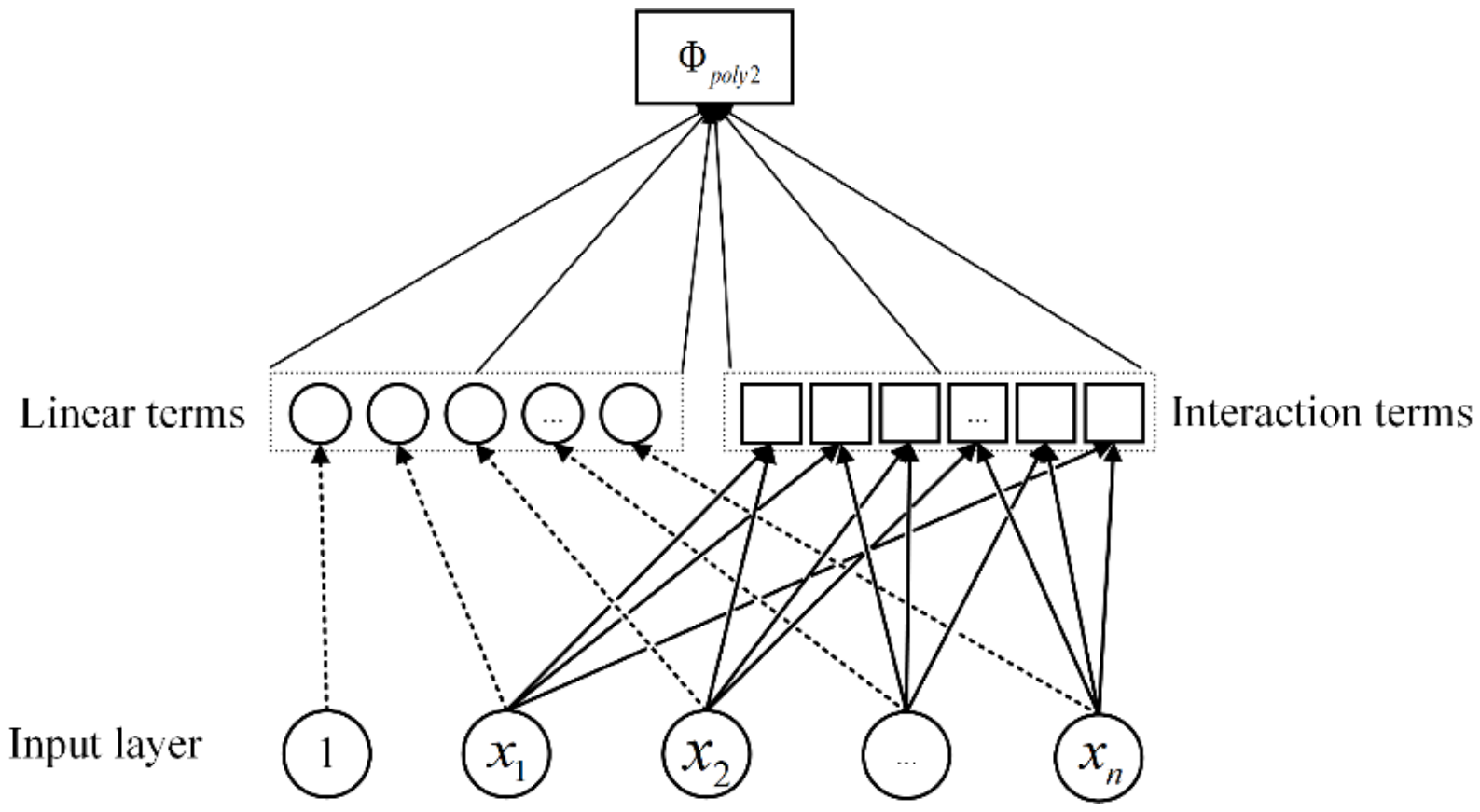

Figure 4. The Poly2 modeling framework.

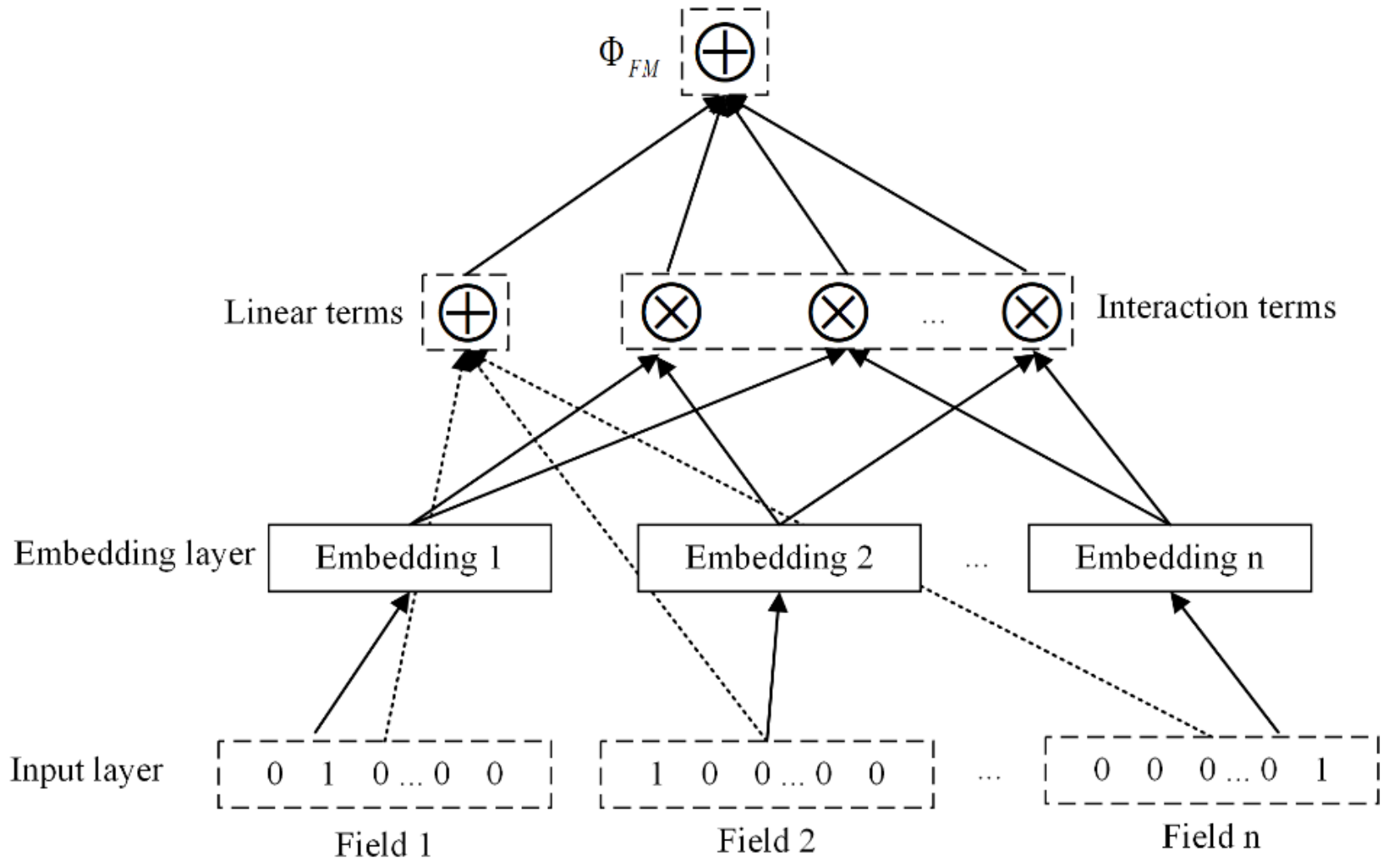

Figure 5. The FMs modeling framework.

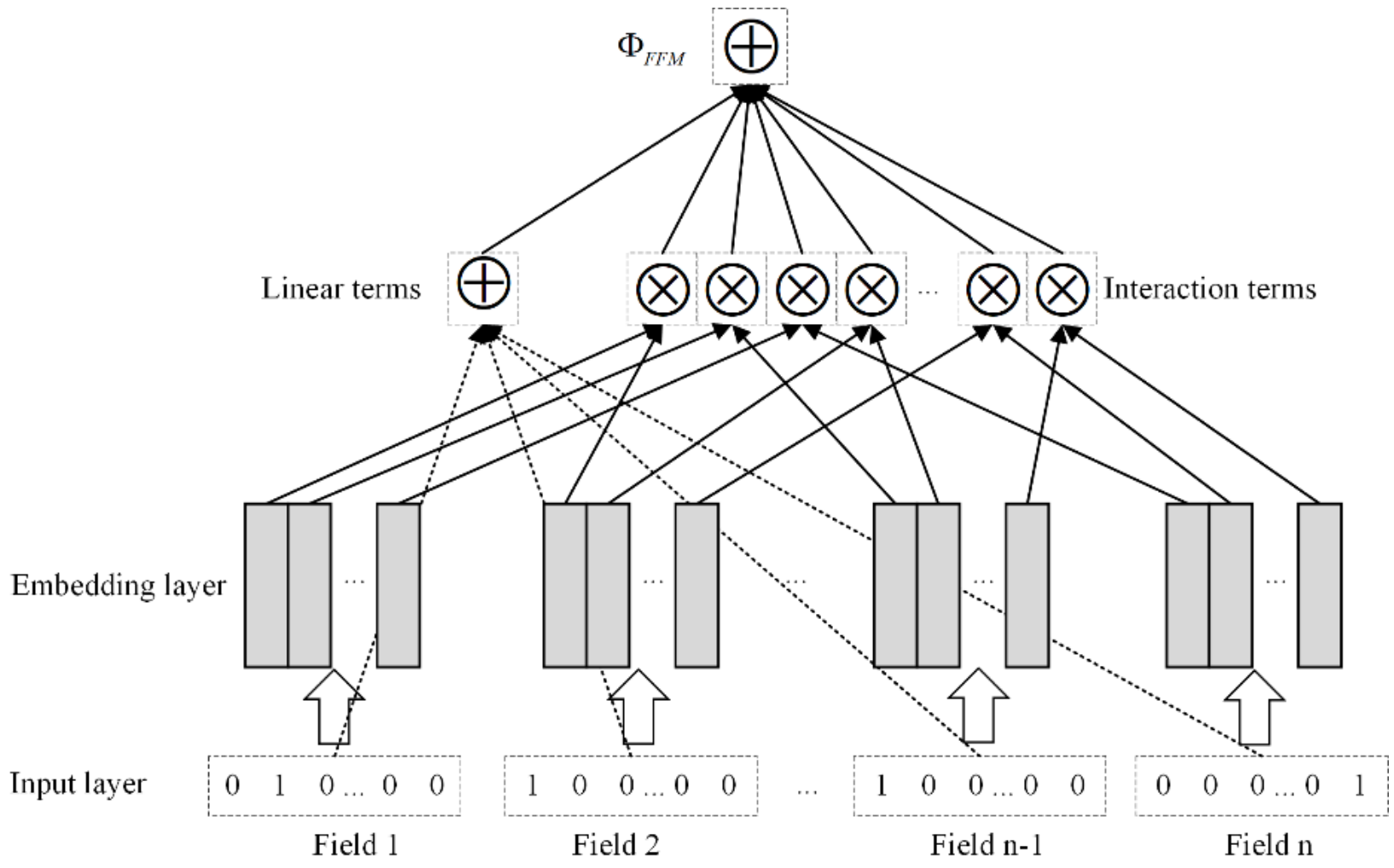

Figure 6. The FFMs modeling framework.

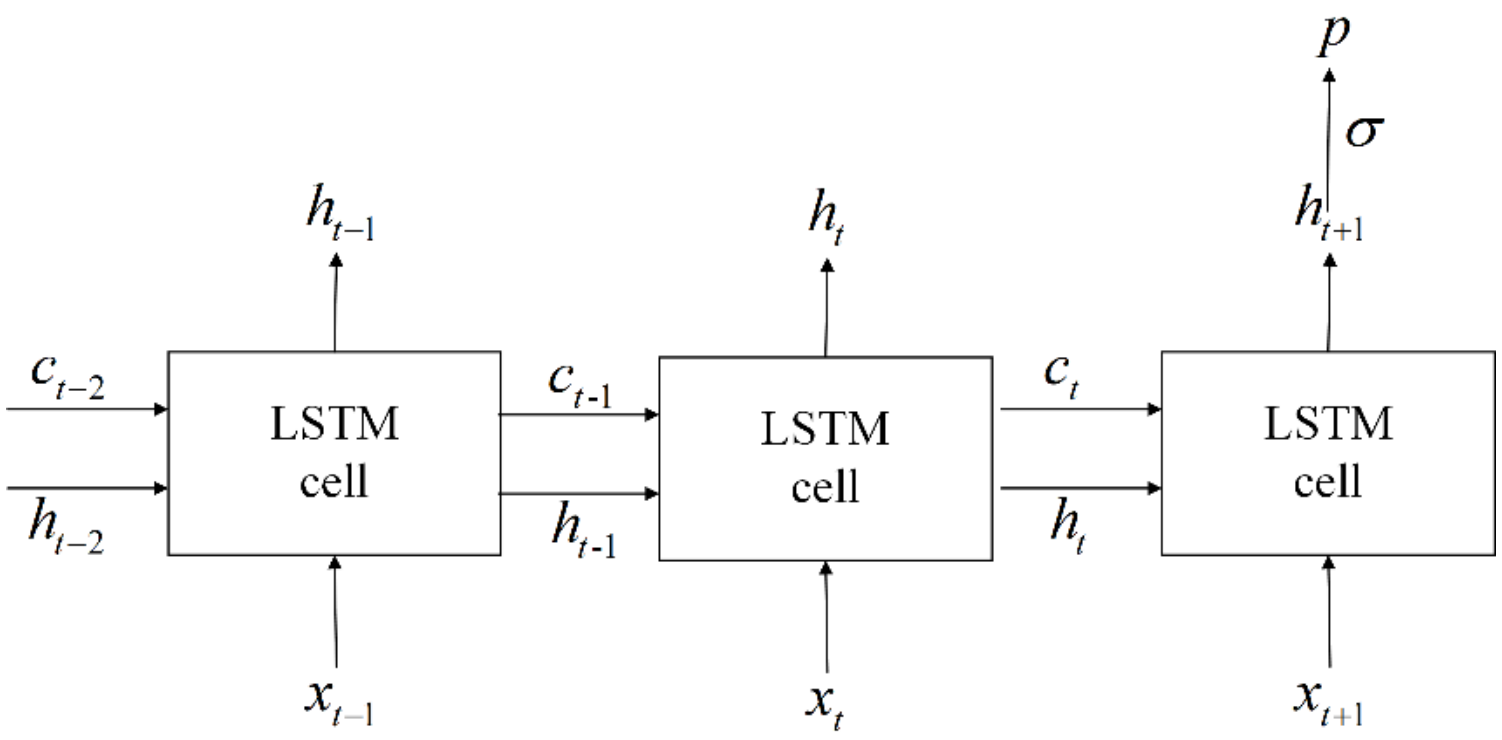

Figure 7. The LSTM modeling framework.

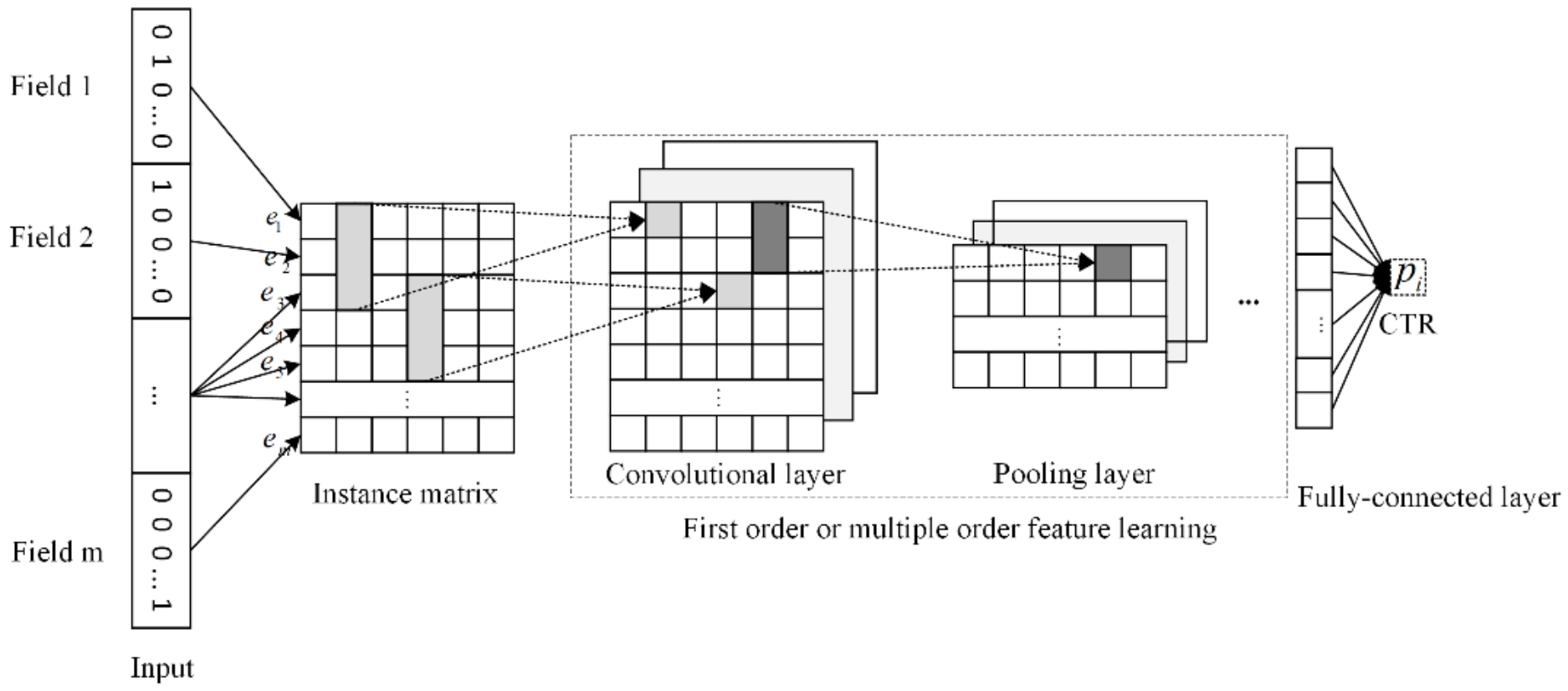

Figure 8. The CNN modeling framework.

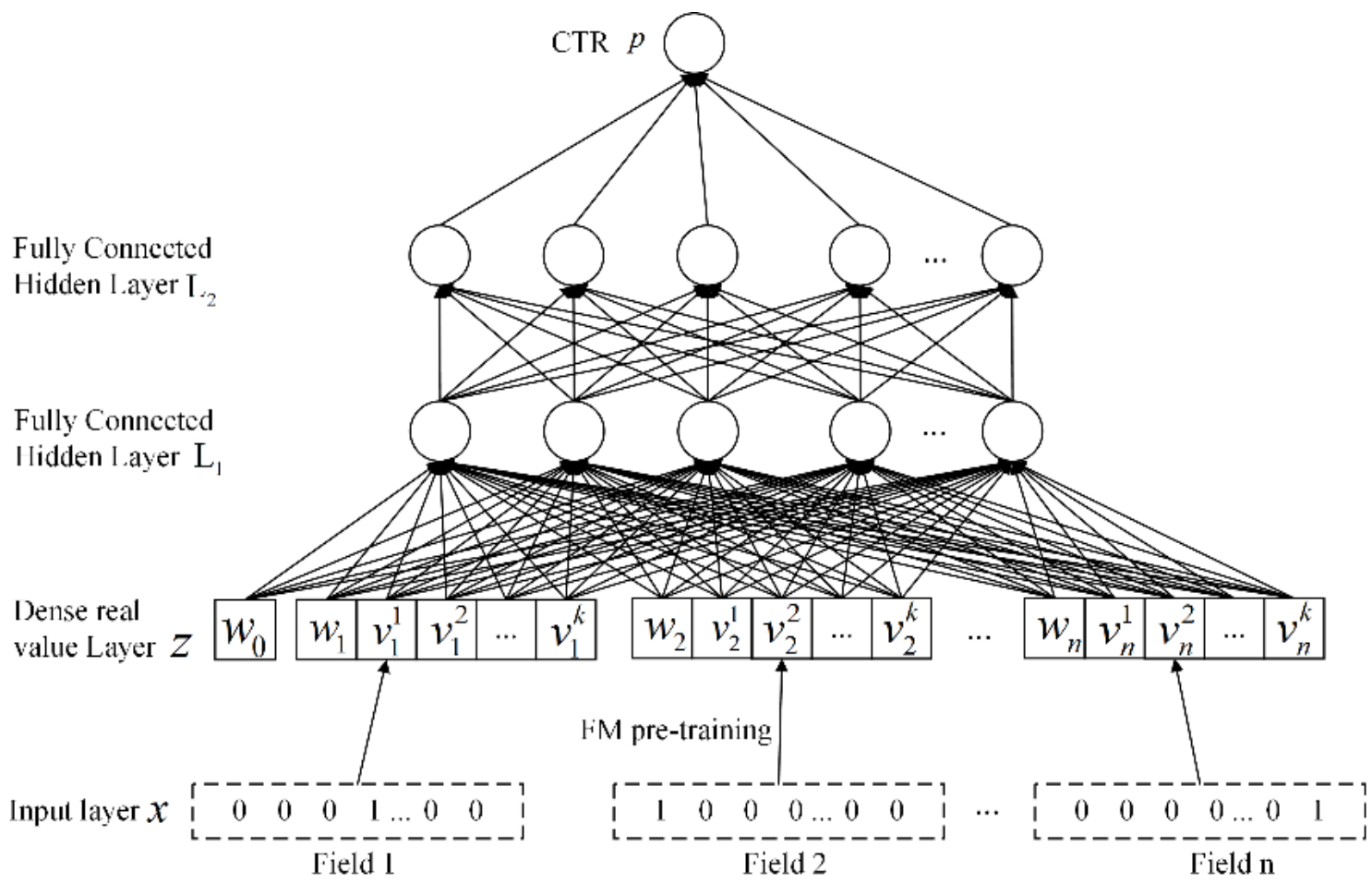

Figure 9. The FNN modeling framework.

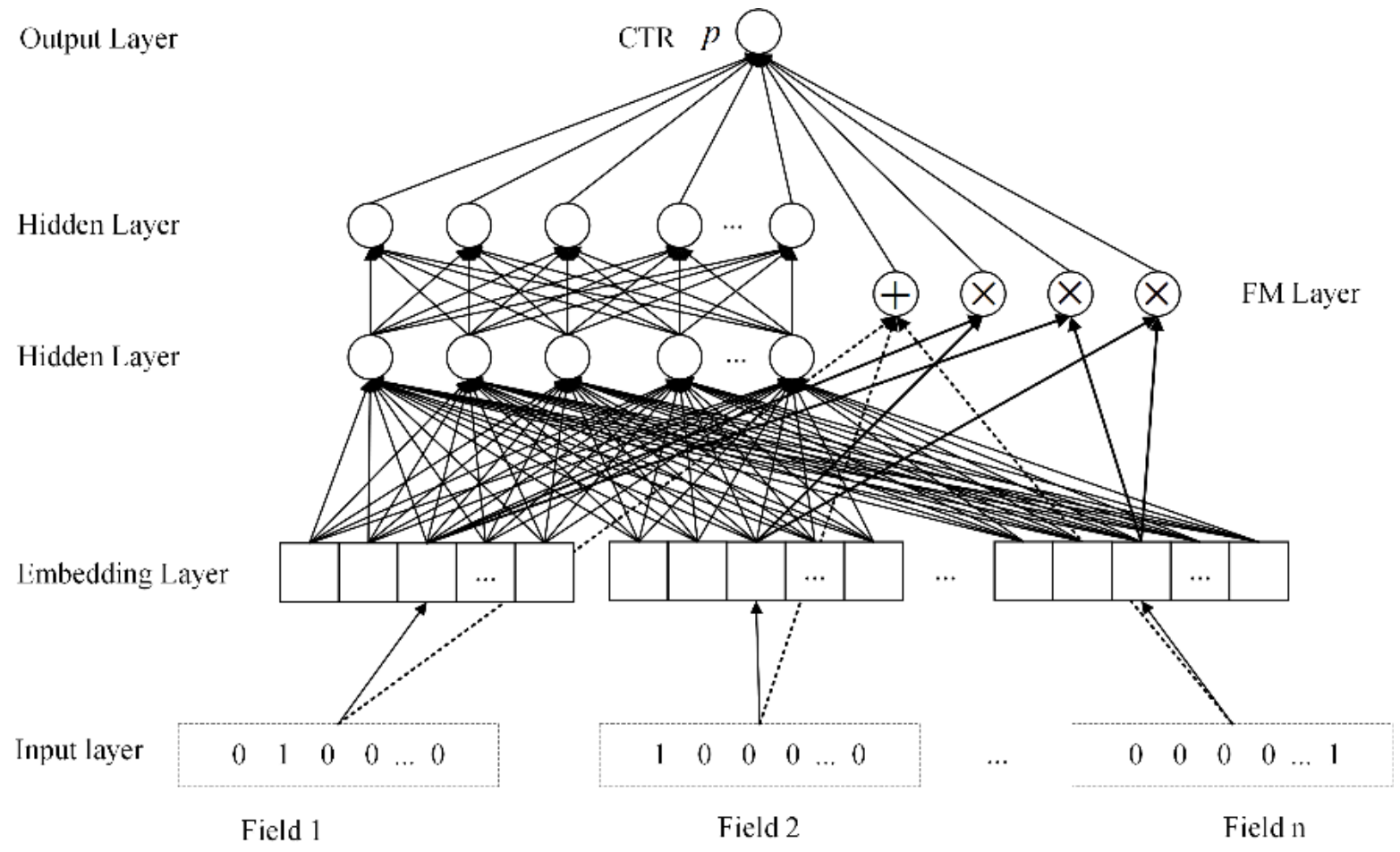

Figure 10. The DeepFM modeling framework.

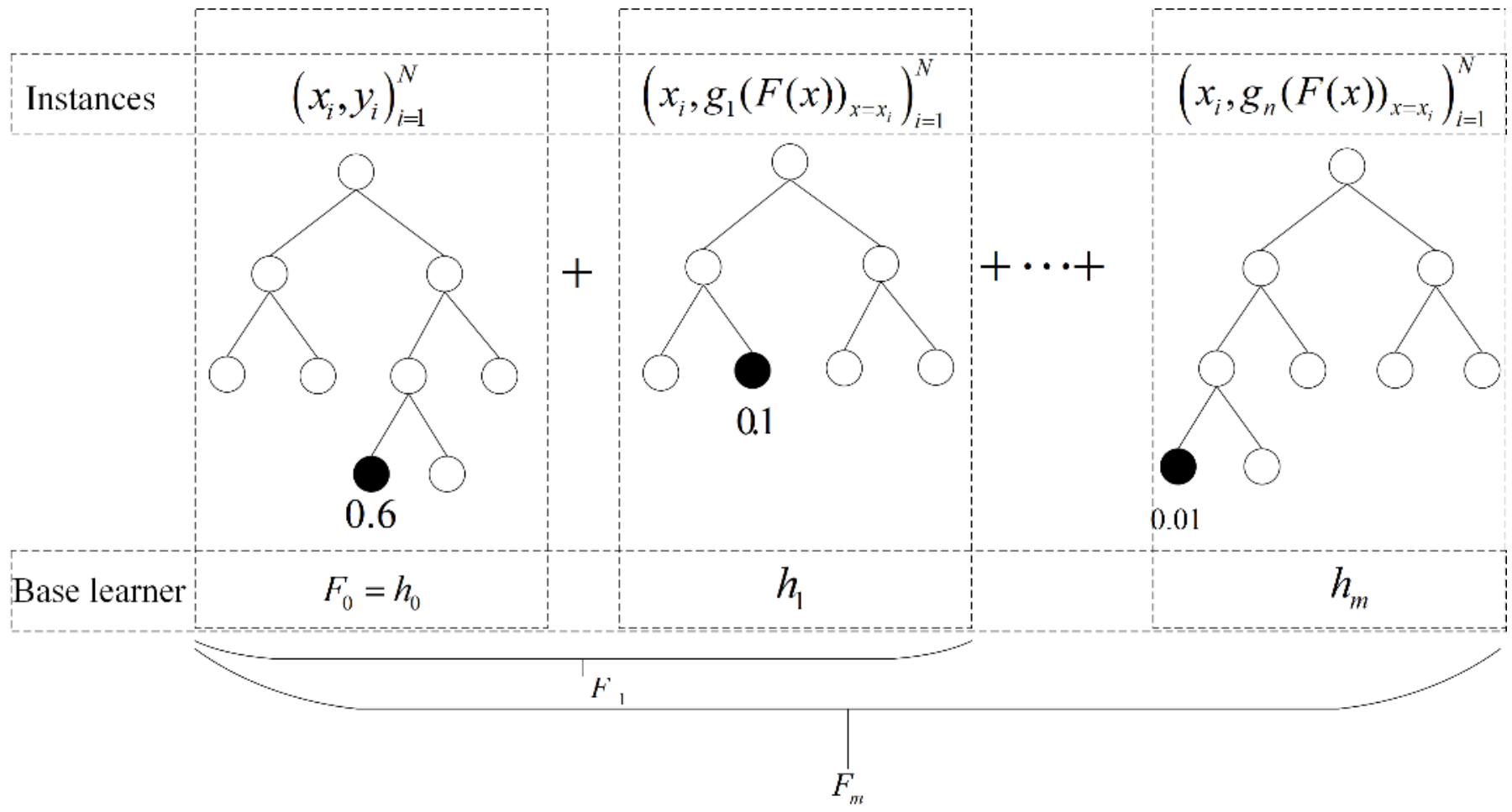

Figure 11. The GBDT modeling framework.

# Appendix for "Click-Through Rate Prediction in Online Advertising: A Literature Review"

Yanwu Yang[1], Panyu Zhai[1]
[1]School of Management, Huazhong University of Science and Technology, Wuhan, China
{yangyanwu.isec,zhaipanyu.isec}@gmail.com

## A.1 Web Search, Recommender System and Online Advertising

Web search, recommender system and online advertising are three most critical information-seeking mechanisms to mitigate the information overload problem (Zhao et al., 2019). On one hand, the three mechanisms share a common procedure of matching users' explicit or implicit needs to a collection of information objects (e.g., Web pages, product descriptions and online advertisements) (Garcia-Molina et al., 2011). From an engineering viewpoint, they are with similar data processing processes and systems architectures.

On the other hand, search, recommender system and advertising differ in the following aspects. (a) Inputs: search takes as the input users' queries that explicitly describe their interests, while recommender system and advertising usually extract users' implicit interests from their behaviors and contexts (Zhang et al., 2020). (b) Outputs: search presents objects that match well users' queries, recommender system generates a set of products (or items) that match users' implicit preferences, and advertising suggests the right advertisements to the right users according to their profiles, behaviors and contexts. In other words, search offers non-commercial and organic results, recommender system provides product descriptions, and advertising handles marketing communications and advertising information (i.e., sponsored results). In contrast to search, advertising is analogous to recommender system in that both dealing with commercial information by analyzing contexts (e.g., what users have browsed and purchased recently) and users' profiles and behaviors (Garcia-Molina et al., 2011). (c) Goals: the main goal of search is to retrieve more relevant and diverse items and present search results in accessible and understandable ways, recommender system aims to provide personalized services by suggesting a set of product items that are likely interesting for users, and advertising is to maximize CTR, conversion rate, and the expected revenue or return on investment (ROI) yielded from advertising campaigns.

Owing to differences in inputs, outputs and goals, as discussed above, CTR prediction in online advertising is distinct from that in search and recommender system, although all the three online mechanisms may pursue maximized clicks and CTR. Specifically, advertising CTR prediction focuses on processing features related to advertising objects and contexts to learn users' clicking behaviors toward advertisements. Moreover, there are a variety of online advertising forms with inherent characteristics (e.g., position, quality and slogan) that may influence the click probability.

## A.2 The Procedure of CTR Prediction

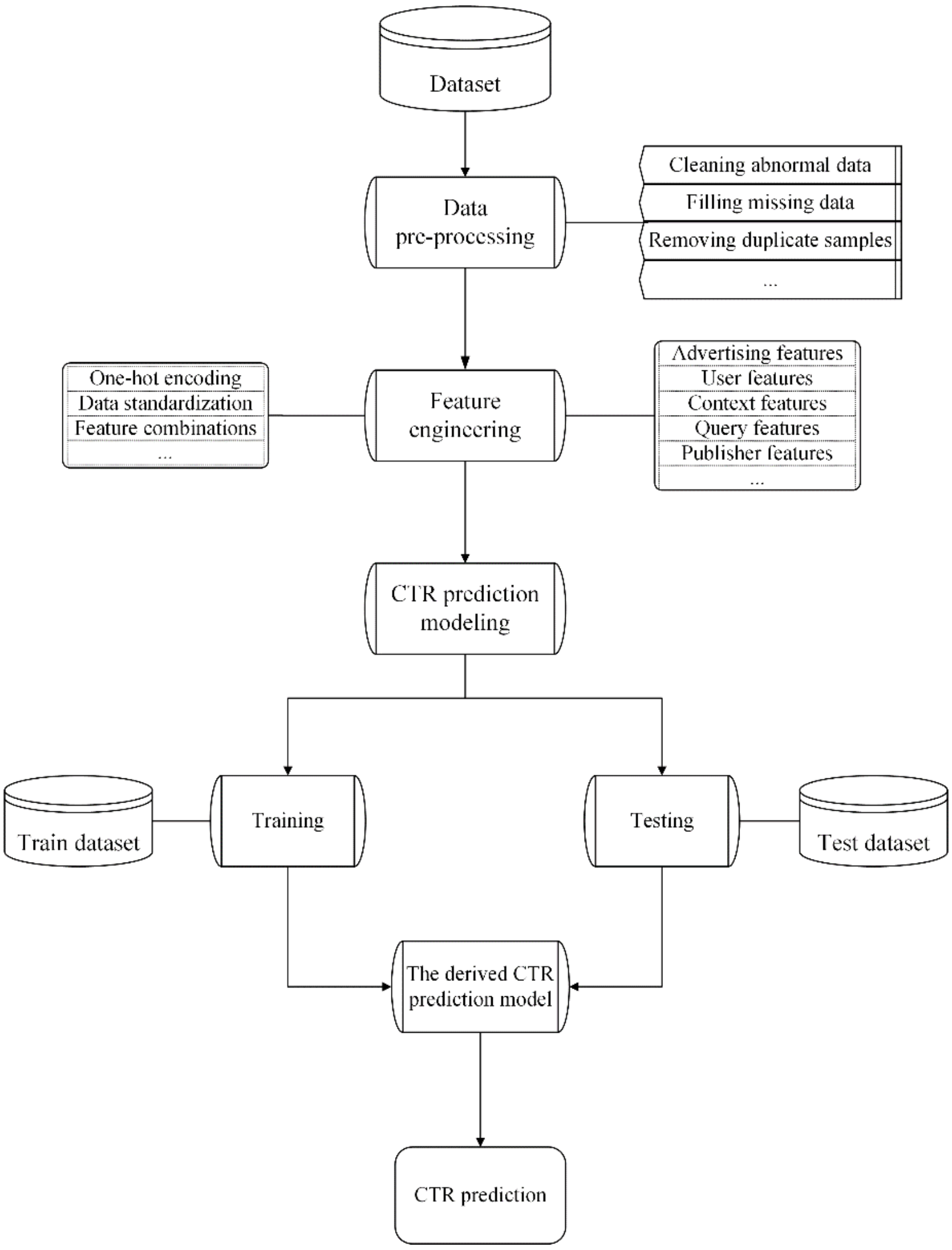

Figure A1. The CTR prediction procedure.

The procedure for CTR prediction includes five steps, as shown in Figure A1. The first step is to pre-process the raw dataset, e.g., cleaning abnormal data, filling missing data, and removing duplicate samples, etc. The second step is to conduct feature engineering based on attributes of the raw data, in order to select and extract useful features for CTR prediction (Qu et al., 2019; Moneera et al., 2021). The third step is to build the CTR prediction model by using statistical or machine learning techniques, which will be elaborated in more detail in Section 4. In the fourth step, the dataset is divided into train subset and test subset: on the former parameters of the CTR prediction model can be obtained through some type of optimization algorithms such as Stochastic Gradient Descent, and on the latter the performance of the trained CTR prediction model is validated. Finally, the CTR prediction model derived from the fourth step is used to compute click probabilities on a new set of samples.

## A.3 Research Articles included in This Review

Table A1. The List of Reviewed Articles.

| No. | Title | Year | Authors | Context | Publication Outlet |
|---|---|---|---|---|---|
| 1 | Predicting clicks: Estimating the click-through rate for new ads | 2007 | Richardson et al. | Search advertising | *The 16th International Conference on World Wide Web (WWW'07)* |
| 2 | Contextual advertising by combining relevance with click feedback | 2008 | Chakrabarti et al. | Contextual advertising | *The 17th international conference on World Wide Web (WWW'08)* |
| 3 | Predicting ads click through rate with decision rules | 2008 | Dembczynski et al. | Search advertising | *Workshop on Targeting and Ranking in Online Advertising* |
| 4 | Temporal click model for sponsored search | 2010 | Xu et al. | Search advertising | *The 33rd International ACM SIGIR Conference on Research and Development in Information Retrieval (SIGIR'10)* |
| 5 | A novel click model and its applications to online advertising | 2010 | Zhu et al. | Search advertising | *The Third ACM International Conference on Web Search and Data Mining (WSDM'10)* |
| 6 | Learning User Behaviors for Advertisements Click Prediction | 2011 | Wang and Chen | Search advertising | *The 34rd international ACM SIGIR conference on research and development in information retrieval Workshop on Internet Advertising (SIGIR'11)* |
| 7 | Multimedia features for click prediction of new ads in display advertising | 2012 | Cheng et al. | Display advertising | *The 18th ACM SIGKDD International Conference on Knowledge Discovery and Data Mining (KDD'12)* |
| 8 | Social Network and Click-through Prediction with Factorization Machines | 2012 | Rendle. | Search advertising | *KDD Cup 2012* |
| 9 | Using boosted trees for click-through rate prediction for sponsored search | 2012 | Trofimov et al. | Search advertising | *The Sixth International Workshop on Data Mining for Online Advertising and Internet Economy (ADKDD'12)* |
| 10 | Click-Through Prediction for Sponsored Search Advertising with Hybrid Models | 2012 | Wang et al. | Search advertising | *KDD Cup 2012* |
| 11 | Relational click prediction for sponsored search | 2012 | Xiong et al. | Search advertising | *The fifth ACM international conference on Web search and data mining (WSDM'12)* |
| 12 | Ad click prediction: A view from the trenches | 2013 | McMahan et al. | Search advertising | *The 19th ACM SIGKDD International Conference on Knowledge Discovery and Data Mining (KDD'13)* |
| 13 | Advertisement Click-Through Rate Prediction Using Multiple Criteria Linear Programming Regression Model | 2013 | Wang et al. | Search advertising | *Procedia Computer Science* |
| 14 | Simple and Scalable Response Prediction for Display Advertising | 2014 | Chapelle et al. | Display advertising | *ACM Transactions on Intelligent Systems and Technology* |
| 15 | Sampling dilemma: Towards effective data sampling for click prediction in sponsored search | 2014 | Feng et al. | Search advertising | *The 7th ACM International Conference on Web Search and Data Mining (WSDM'14)* |
| 16 | Practical Lessons from Predicting Clicks on Ads at Facebook | 2014 | He et al. | Social media advertising | *The 20th ACM SIGKDD Conference on Knowledge Discovery and Data Mining (ADKDD'14)* |
| 17 | IPinYou Global RTB Bidding Algorithm Competition Dataset | 2014 | Liao et al. | Display advertising | *The 20th ACM SIGKDD Conference on Knowledge Discovery and Data Mining (ADKDD'14)* |
| 18 | Coupled Group Lasso for Web-Scale CTR Prediction in Display Advertising | 2014 | Yan et al. | Display advertising | *The 31st International Conference on Machine Learning (ICML'14)* |
| 19 | Real-Time Bidding Benchmarking with iPinYou Dataset | 2014 | Zhang et al. | Display advertising | *Pre-print (arXiv)* |
| 20 | Sequential Click Prediction for Sponsored Search with Recurrent Neural Networks | 2014 | Zhang et al. | Search advertising | *The AAAI Conference on Artificial Intelligence (AAAI' 14)* |
| 21 | Predicting clicks: CTR estimation of advertisements using Logistic Regression classifier | 2015 | Kumar et al. | Search advertising | *The 2015 IEEE International Advance Computing Conference (IACC)* |
| 22 | Click-through Prediction for Advertising in Twitter Timeline | 2015 | Li et al. | Social media advertising | *The 21th ACM SIGKDD International Conference on Knowledge Discovery and Data Mining (KDD15)* |
| 23 | A Convolutional Click Prediction Model | 2015 | Liu et al. | Mobile advertising | *The 24th ACM International on Conference on Information and Knowledge Management (CIKM'15)* |
| 24 | Factorization machines with follow-the-regularized-leader for CTR prediction in display advertising | 2015 | Ta | Display advertising | *The 2015 IEEE International Conference on Big Data (Big Data)* |
| 25 | Click Through Rate Prediction for Display Advertisement | 2016 | Avila Clemenshia & Vijaya | Display advertising | *International Journal of Computer Applications* |
| 26 | XGBoost: A Scalable Tree Boosting System | 2016 | Chen & Guestrin | Display advertising | *The 22nd ACM SIGKDD International Conference on Knowledge Discovery and Data Mining* |
| 27 | Deep CTR Prediction in Display Advertising | 2016 | Chen et al. | Display advertising | *The 24th ACM International Conference on Multimedia (MM'16)* |
| 28 | Estimating Ads' Click through Rate with Recurrent Neural Network | 2016 | Chen et al. | Mobile advertising | *The 3rd Annual International Conference on Information Technology and Applications (ITA 2016)* |
| 29 | Research on CTR Prediction for Contextual Advertising Based on Deep Architecture Model | 2016 | Jiang et al. | Contextual advertising | *Control Engineering and Applied Informatics* |
| 30 | Field-aware Factorization Machines for CTR Prediction | 2016 | Juan et al. | Display advertising | *The 10th ACM Conference on Recommender Systems (RecSys '16)* |
| 31 | F2M: Scalable Field-Aware Factorization Machines | 2016 | Ma et al. | Display advertising | *The 30th Conference on Neural Information Processing Systems (NIPS 2016)* |
| 32 | A Communication-Efficient Parallel Algorithm for Decision Tree | 2016 | Meng et al. | Search advertising | *Pre-print (arXiv)* |
| 33 | Sparse Factorization Machines for Click-through Rate Prediction | 2016 | Pan et al. | Display advertising | *The 16th International Conference on Data Mining (ICDM2016)* |
| 34 | Product-based Neural Networks for User Response Prediction | 2016 | Qu et al. | Display advertising | *The 2016 IEEE 16th International Conference on Data Mining (ICDM)* |
| 35 | Predicting ad click-through rates via feature-based fully coupled interaction tensor factorization | 2016 | Shan et al. | Display advertising | *Electronic Commerce Research and Applications* |
| 36 | Feature Selection in Click-Through Rate Prediction Based on Gradient Boosting | 2016 | Wang et al. | Display advertising | *Intelligent Data Engineering and Automated Learning (IDEAL)* |
| 37 | Deep Learning over Multi-field Categorical Data - A Case Study on User Response Prediction | 2016 | Zhang et al. | Display advertising | *European conference on information retrieval* |
| 38 | Deep Character-Level Click-Through Rate Prediction for Sponsored Search | 2017 | Edizel et al. | Search advertising | *The 40th International ACM SIGIR Conference on Research and Development in Information Retrieval (SIGIR '17)* |
| 39 | DeepFM: A Factorization-Machine based Neural Network for CTR Prediction | 2017 | Guo et al. | Display advertising | *Pre-print (arXiv)* |
| 40 | An Ad CTR Prediction Method Based on Feature Learning of Deep and Shallow Layers | 2017 | Huang et al. | Mobile advertising | *The 2017 ACM Conference on Information and Knowledge Management (CIKM'17)* |
| 41 | A CTR Prediction Approach for Text Advertising Based on the SAE-LR Deep Neural Network | 2017 | Jiang et al. | Search advertising | *Journal of Information Processing Systems* |
| 42 | A CTR prediction method based on feature engineering and online learning | 2017 | Jie-Hao et al. | Search advertising | *The 17th International Symposium on Communications and Information Technologies (ISCIT2017)* |
| 43 | Field-aware Factorization Machines in a Real-world Online Advertising System | 2017 | Juan et al. | Display Advertising | *The 26th International Conference on World Wide Web Companion (WWW '17)* |
| 44 | LightGBM: A Highly Efficient Gradient Boosting Decision Tree | 2017 | Ke et al. | Search advertising | *Advances in neural information processing systems* |
| 45 | Model Ensemble for Click Prediction in Bing Search Ads | 2017 | Ling et al. | Search advertising | *The 26th International Conference on World Wide Web Companion (WWW '17)* |
| 46 | PBODL: Parallel Bayesian Online Deep Learning for Click-Through Rate Prediction in Tencent Advertising System | 2017 | Liu et al. | Social media advertising, mobile advertising | *Pre-print (arXiv)* |
| 47 | Deep & Cross Network for Ad Click Predictions | 2017 | Wang et al. | Display advertising | *The ADKDD'17(ADKDD'17)* |
| 48 | Advertisement Click-Through Rate Prediction Based on the Weighted-ELM and Adaboost Algorithm | 2017 | Zhang et al. | Display advertising | *Scientific Programming* |
| 49 | Convolutional Neural Networks based Click-Through Rate Prediction with Multiple Feature Sequences | 2018 | Chan et al. | Social media advertising, mobile advertising | *The Twenty-Seventh International Joint Conference on Artificial Intelligence (IJCAI '18)* |
| 50 | Ad Click Prediction in Sequence with Long Short-Term Memory Networks: An Externality-aware Model | 2018 | Deng et al. | Search advertising | *The 41st International ACM SIGIR Conference on Research & Development in Information Retrieval (SIGIR '18)* |
| 51 | Wide & ResNet: An Improved Network for CTR Prediction | 2018 | Gao & Bie | Display advertising | *The 2018 International Conference on Algorithms, Computing and Artificial Intelligence (ACAI 2018)* |
| 52 | DeepFM: An End-to-End Wide & Deep Learning Framework for CTR Prediction | 2018 | Guo et al. | Display advertising | *Pre-print (arXiv)* |
| 53 | A CTR Prediction Approach for Advertising Based on Embedding Model and Deep Learning | 2018 | Jiang et al. | Contextual Advertising | *The 2018 IEEE Intl Conf on Parallel & Distributed Processing with Applications, Ubiquitous Computing & Communications, Big Data & Cloud Computing, Social Computing & Networking, Sustainable Computing& Communications* |
| 54 | xDeepFM: Combining Explicit and Implicit Feature Interactions for Recommender Systems | 2018 | Lian et al. | Display advertising | *The 24th ACM SIGKDD International Conference on Knowledge Discovery & Data Mining* |
| 55 | Field-aware probabilistic embedding neural network for CTR prediction | 2018 | Liu et al. | Display advertising, mobile advertising | *The 12th ACM Conference on Recommender Systems (RecSys'18)* |
| 56 | Field-weighted Factorization Machines for Click-Through Rate Prediction in Display Advertising | 2018 | Pan et al. | Display advertising | *The 2018 World Wide Web Conference (WWW '18)* |
| 57 | Optimizing Field-Aware Factorization Machine with Particle Swarm Optimization on Online Ads Click-through Rate Prediction | 2018 | Phangtriastu & Isa | Display advertising | *The 2018 3rd International Conference on Computer and Communication Systems (ICCCS)* |
| 58 | ETCF: An Ensemble Model for CTR Prediction | 2018 | Qiu et al. | Display advertising, mobile advertising | *The 2018 15th International Conference on Service Systems and Service Management (ICSSSM)* |
| 59 | Research on Advertising Click-Through Rate Prediction Based on CNN-FM Hybrid Model | 2018 | She & Wang | Mobile advertising | *The 2018 10th International Conference on Intelligent Human-Machine Systems and Cybernetics (IHMSC)* |
| 60 | Click-through Rate Estimates based on Deep Learning | 2018 | Wang & He | Search advertising | *The 2018 2nd International Conference on Deep Learning Technologies (ICDLT '18)* |
| 61 | A New Approach for Advertising CTR Prediction Based on Deep Neural Network via Attention Mechanism | 2018 | Wang et al. | Search advertising | *Computational and Mathematical Methods in Medicine* |
| 62 | Deep Interest Network for Click-Through Rate Prediction | 2018 | Zhou et al. | Display advertising | *The 24th ACM SIGKDD International Conference on Knowledge Discovery & Data Mining (KDD '18)* |
| 63 | FLEN: Leveraging Field for Scalable CTR Prediction | 2019 | Chen et al. | Mobile advertising | *Pre-print (arXiv)* |
| 64 | Deeply supervised model for click-through rate prediction in sponsored search | 2019 | Gligorijevic et al. | Search advertising | *Data Mining and Knowledge Discovery* |
| 65 | Order-aware Embedding Neural Network for CTR Prediction | 2019 | Guo et al. | Display advertising, mobile advertising | *The 42nd International ACM SIGIR Conference on Research and Development in Information Retrieval (SIGIR '19)* |
| 66 | FiBiNET: Combining feature importance and bilinear feature interaction for click-through rate prediction | 2019 | Huang et al. | Display advertising, mobile advertising | *The 13th ACM Conference on Recommender Systems (RecSys '19)* |
| 67 | DeepGBM: A Deep Learning Framework Distilled by GBDT for Online Prediction Tasks | 2019 | Ke et al. | Display advertising | *The 25th ACM SIGKDD International Conference on Knowledge Discovery & Data Mining (KDD '19)* |
| 68 | Fi-GNN: Modeling Feature Interactions via Graph Neural Networks for CTR Prediction | 2019 | Li et al. | Display advertising & Mobile advertising | *The 28th ACM International Conference on Information and Knowledge Management (CIKM '19)* |
| 69 | Feature Generation by Convolutional Neural Network for Click-Through Rate Prediction | 2019 | Liu et al. | Display advertising, mobile advertising | *The World Wide Web Conference on (WWW '19)* |
| 70 | AutoCross: Automatic Feature Crossing for Tabular Data in Real-World Applications | 2019 | Luo et al. | Display advertising | *The 25th ACM SIGKDD International Conference on Knowledge Discovery & Data Mining (KDD '19)* |
| 71 | Click-through rate prediction with the user memory network | 2019 | Ouyang et al. | Context advertising, search advertising | *The 1st International Workshop on Deep Learning Practice for High-Dimensional Sparse Data (DLP-KDD'19)* |
| 72 | Deep Spatio-Temporal Neural Networks for Click-Through Rate Prediction | 2019 | Ouyang et al. | Contextual advertising, search advertising | *The 25th ACM SIGKDD International Conference on Knowledge Discovery & Data Mining (KDD '19)* |
| 73 | Representation Learning-Assisted Click-Through Rate Prediction. | 2019 | Ouyang et al. | Context advertising, search advertising | *Pre-print (arXiv)* |
| 74 | Warm Up Cold-start Advertisements: Improving CTR Predictions via Learning to Learn ID Embeddings | 2019 | Pan et al. | Search advertising | *The 42nd International ACM SIGIR Conference on Research and Development in Information Retrieval (SIGIR '19)* |
| 75 | Practice on Long Sequential User Behavior Modeling for Click-Through Rate Prediction | 2019 | Pi et al. | Display advertising | *The 25th ACM SIGKDD International Conference on Knowledge Discovery & Data Mining (KDD '19)* |
| 76 | A Dynamic Neural Network Model for Click-Through Rate Prediction in Real-Time Bidding | 2019 | Qu et al. | Display advertising | *The 2019 IEEE International Conference on Big Data (Big Data)* |
| 77 | An Embedded Model XG-FwFMs for Click-Through Rate | 2019 | Shi et al. | Display advertising | *The 2019 4th International Conference on Big Data and Computing (ICBDC 2019)* |
| 78 | Research on CTR prediction based on stacked autoencoder | 2019 | Wang et al. | Search advertising | *Applied Intelligence* |
| 79 | Learning over categorical data using counting features: With an application on click-through rate estimation | 2019 | Wu et al. | Display Advertising | *The 1st International Workshop on Deep Learning Practice for High-Dimensional Sparse Data (DLP-KDD '19)* |
| 80 | Robust transfer integrated locally kernel embedding for click-through rate prediction | 2019 | Xie et al. | Display advertising, mobile advertising, context advertising | *Information Sciences* |
| 81 | Improving Ad Click Prediction by Considering Non-displayed Events | 2019 | Yuan et al. | Search advertising | *The 28th ACM International Conference on Information and Knowledge Management (CIKM '19)* |
| 82 | Field-Aware Neural Factorization Machine for Click-Through Rate Prediction | 2019 | Zhang et al. | Mobile advertising | *IEEE Access* |
| 83 | Deep interest evolution network for click-through rate prediction | 2019 | Zhou et al. | Display advertising | *The AAAI conference on artificial intelligence (AAAI-19)* |
| 84 | XGBDeepFM for CTR Predictions in Mobile Advertising Benefits from Ad Context | 2020 | An & Ren | Mobile advertising | *Mathematical Problems in Engineering* |
| 85 | AutoConjunction: Adaptive Model-based Feature Conjunction for CTR Prediction | 2020 | Chang et al. | Display advertising, mobile advertising | *The 2020 21st IEEE International Conference on Mobile Data Management (MDM)* |
| 86 | Metadata Matters in User Engagement Prediction | 2020 | Chen et al. | Display image advertising | *The 43rd International ACM SIGIR Conference on Research and Development in Information Retrieval (SIGIR '20).* |
| 87 | Differentiable neural input search for recommender systems | 2020 | Cheng et al. | Display advertising | *Pre-print (arXiv)* |
| 88 | Deep Learning for User Interest and Response Prediction in Online Display Advertising | 2020 | Gharibshah et al. | Display advertising | *Data Science and Engineering* |
| 89 | A New Click-Through Rates Prediction Model Based on Deep&Cross Network | 2020 | Huang et al. | Display advertising | *Algorithms* |
| 90 | AutoFeature: Searching for Feature Interactions and Their Architectures for Click-through Rate Prediction | 2020 | Khawar et al. | Display advertising & Mobile advertising | *The 29th ACM International Conference on Information & Knowledge Management (CIKM '20)* |
| 91 | A CTR prediction model based on user interest via attention mechanism | 2020 | Li et al. | Display advertising | *Applied Intelligence* |
| 92 | Interpretable Click-Through Rate Prediction through Hierarchical Attention | 2020 | Li et al. | Display advertising, mobile advertising | *The 13th International Conference on Web Search and Data Mining (WSDM '20)* |
| 93 | FINET: Fine-grained Feature Interaction Network for Click-through Rate Prediction | 2020 | Lian & Ge | Display advertising, mobile advertising | *The 2020 12th International Conference on Advanced Computational Intelligence (ICACI)* |
| 94 | AutoFIS: Automatic Feature Interaction Selection in Factorization Models for Click-Through Rate Prediction | 2020 | Liu et al. | Display advertising & Mobile advertising | *The 26th ACM SIGKDD International Conference on Knowledge Discovery & Data Mining (KDD '20)* |
| 95 | Iterative Boosting Deep Neural Networks for Predicting Click-Through Rate | 2020 | Livne et al. | Mobile advertising | *Pre-print (arXiv)* |
| 96 | Network On Network for Tabular Data Classification in Real-world Applications | 2020 | Luo et al. | Display advertising & Mobile advertising | *The 43rd International ACM SIGIR Conference on Research and Development in Information Retrieval (SIGIR '20)* |
| 97 | Density Matrix Based Convolutional Neural Network for Click-Through Rate Prediction | 2020 | Niu & Hou | Display advertising, mobile advertising | *The 2020 3rd International Conference on Artificial Intelligence and Big Data (ICAIBD)* |
| 98 | Field-aware Calibration: A Simple and Empirically Strong Method for Reliable Probabilistic Predictions | 2020 | Pan et al. | Display advertising & Mobile advertising | *The Web Conference 2020 (WWW '20)* |
| 99 | Search-based User Interest Modeling with Lifelong Sequential Behavior Data for Click-Through Rate Prediction | 2020 | Pi et al. | Display advertising | *The 29th ACM International Conference on Information & Knowledge Management* |
| 100 | Feature Engineering of Click-through-rate Prediction for Advertising | 2020 | Ren et al. | Search advertising | *The International Conference in Communications, Signal Processing, and Systems* |
| 101 | HFF: Hybrid Feature Fusion Model for Click-Through Rate Prediction | 2020 | Shi & Yang | Display advertising, mobile advertising, search advertising | *The 2018 10th International Conference on Intelligent Human-Machine Systems and Cybernetics (IHMSC)* |
| 102 | Compositional Embeddings Using Complementary Partitions for Memory-Efficient Recommendation Systems | 2020 | Shi et al. | Display advertising | *The 26th ACM SIGKDD International Conference on Knowledge Discovery & Data Mining (KDD '20)* |
| 103 | Towards Automated Neural Interaction Discovery for Click-Through Rate Prediction | 2020 | Song et al. | Display advertising, mobile advertising, search advertising | *The 26th ACM SIGKDD International Conference on Knowledge Discovery & Data Mining (KDD '20)* |
| 104 | HoAFM: A High-order Attentive Factorization Machine for CTR Prediction | 2020 | Tao et al. | Display advertising, | *Information Processing & Management* |
| | | | | mobile advertising | |
| 105 | A Hierarchical Attention Model for CTR Prediction Based on User Interest | 2020 | Wang et al. | Display advertising | *IEEE Systems Journal* |
| 106 | Attention-over-Attention Field-Aware Factorization Machine. | 2020 | Wang et al. | Display advertising | *The AAAI Conference on Artificial Intelligence* |
| 107 | AutoRec: An Automated Recommender System | 2020 | Wang et al. | Display advertising & Mobile advertising | *The fourteenth ACM Conference on Recommender Systems (RecSys '20)* |
| 108 | TFNet: Multi-Semantic Feature Interaction for CTR Prediction | 2020 | Wu et al. | Display advertising & Mobile advertising | *The 43rd International ACM SIGIR Conference on Research and Development in Information Retrieval (SIGIR '20)* |
| 109 | Modeling low- and high-order feature interactions with FM and self-attention network | 2020 | Yan et al. | Display advertising, mobile advertising | *Applied Intelligence* |
| 110 | Operation-aware Neural Networks for user response prediction | 2020 | Yang et al. | Display advertising | *Neural Networks* |
| 111 | Deep Interaction Machine: A Simple but Effective Model for High-order Feature Interactions | 2020 | Yu et al. | Display advertising & Mobile advertising | *The 29th ACM International Conference on Information & Knowledge Management (CIKM '20)* |
| 112 | Unbiased Ad Click Prediction for Position-aware Advertising Systems | 2020 | Yuan et al. | Display advertising | *The fourteenth ACM Conference on Recommender Systems (RecSys '20)* |
| 113 | An Attention-based Deep Network for CTR Prediction | 2020 | Zhang et al. | Display advertising | *The 12th International Conference on Machine Learning and Computing (ICMLC 2020)* |
| 114 | Amer: Automatic behavior modeling and interaction exploration in recommender system | 2020 | Zhao et al. | Display advertising & Mobile advertising | *Pre-print (arXiv)* |
| 115 | Memory-efficient embedding for recommendations | 2020 | Zhao et al. | Display advertising & Mobile advertising | *Pre-print (arXiv)* |
| 116 | Ensembled CTR Prediction via Knowledge Distillation | 2020 | Zhu et al. | Display advertising & Mobile advertising | *The 29th ACM International Conference on Information & Knowledge Management (CIKM '20)* |
| 117 | Feature Interaction based Neural Network for Click-Through Rate Prediction | 2020 | Zou et al. | Display advertising, mobile advertising | *Pre-print (arXiv)* |
| 118 | An Embedding Learning Framework for Numerical Features in CTR Prediction | 2021 | Guo et al. | Display advertising | *The 27th ACM SIGKDD Conference on Knowledge Discovery & Data Mining (KDD '21)* |
| 119 | Dual Graph enhanced Embedding Neural Network for CTR Prediction | 2021 | Guo et al. | Display advertising | *The 27th ACM SIGKDD Conference on Knowledge Discovery & Data Mining (KDD '21)* |
| 120 | ScaleFreeCTR: MixCache-based Distributed Training System for CTR Models with Huge Embedding Table | 2021 | Guo et al. | Display advertising | *The 44th International ACM SIGIR Conference on Research and Development in Information Retrieval (SIGIR '21)* |
| 121 | Training Recommender Systems at Scale: Communication-Efficient Model and Data Parallelism | 2021 | Gupta et al. | Display advertising | *The 27th ACM SIGKDD Conference on Knowledge Discovery & Data Mining (KDD '21)* |
| 122 | Deep Position-wise Interaction Network for CTR Prediction | 2021 | Huang et al. | Ssearch advertising | *The 44th International ACM SIGIR Conference on Research and Development in Information Retrieval (SIGIR '21)* |
| 123 | Multi-view feature transfer for click-through rate prediction | 2021 | Jiang et al. | Mobile advertising, context advertising | *Information Sciences* |
| 124 | Attentive capsule network for click-through rate and conversion rate prediction in online advertising | 2021 | Li et al. | Display advertising | *Knowledge-Based Systems* |
| 125 | GraphFM: Graph Factorization Machines for Feature Interaction Modeling | 2021 | Li et al. | Display advertising & Mobile advertising | *Pre-print (arXiv)* |
| 126 | Learnable Embedding Sizes for Recommender Systems | 2021 | Liu et al. | Display advertising & Mobile advertising | *The 9th International Conference on Learning Representations (ICLR 2021)* |
| 127 | A General Method For Automatic Discovery of Powerful Interactions In Click-Through Rate Prediction | 2021 | Meng et al. | Display advertising | *The 44th International ACM SIGIR Conference on Research and Development in Information Retrieval (SIGIR '21)* |
| | | | | & Mobile advertising | |
| 128 | Click Through Rate Effectiveness Prediction on Mobile Ads Using Extreme Gradient Boosting | 2021 | Moneera et al. | Mobile advertising | *CMC-Computers Materials & Continua* |
| 129 | Learning Graph Meta Embeddings for Cold-Start Ads in Click-Through Rate Prediction | 2021 | Ouyang et al. | Search advertising, news feed advertising | *The 44th International ACM SIGIR Conference on Research and Development in Information Retrieval (SIGIR '21)* |
| 130 | FM2: Field-matrixed Factorization Machines for Recommender Systems | 2021 | Sun et al. | Display advertising, mobile advertising | *The Web Conference 2021 (WWW'21)* |
| 131 | FIVES: Feature Interaction Via Edge Search for Large-Scale Tabular Data | 2021 | Xie et al. | Display advertising | *The 27th ACM SIGKDD Conference on Knowledge Discovery & Data Mining (KDD '21)* |
| 132 | RLNF: Reinforcement Learning based Noise Filtering for Click-Through Rate Prediction | 2021 | Zhao et al. | Mobile advertising & search advertising | *The 44th International ACM SIGIR Conference on Research and Development in Information Retrieval (SIGIR '21)* |

## A.4 Datasets for Advertising CTR Prediction

Table A2. The Summary of Datasets for Advertising CTR Prediction.

| **Dataset** | **URL** | **Descriptions** | **References** |
|---|---|---|---|
| Criteo-Kaggle display advertising challenge 2014 | https://www.kaggle.com/c/criteo-display-ad-challenge | The Cerito display advertising challenge dataset was provided by CriteoLab on Kaggle in 2014. The dataset was published in Criteo Display Advertising Challenge that is a small subset of Criteo. This dataset includes 13 numerical fields and 26 categorical fields. The purpose of the dataset is to predict which ad will be clicked given a user. | Chapelle et al. (2014); Chen & Guestrin (2016); Wang et al. (2016); Ma et al. (2016); Juan et al. (2017); Guo et al. (2017); Wang et al. (2017); Guo et al. (2018); Pan et al. (2018); Lian et al. (2018); Qiu et al. (2018); Phangtriastu & Isa (2018); Gao & Bie (2018); Ke et al. (2019); Luo et al. (2019); Li et al. (2019); Li et al. (2020a); Chang et al. (2020); Tao et al. (2020); Wang et al. (2020c); Lian & Ge (2020); Shi & Yang (2020); Li et al. (2020b); Yan et al. (2020); Pan et al. (2020); Song et al. (2020); Shi et al. (2020); Yang et al. (2020); Wang et al. (2020b); Yu et al. (2020); Zhao et al. (2020b); Zhu et al. (2020); Sun et al. (2021); Guo et al. (2021a); Gupta et al. (2021); Li et al. (2021b); Liu et al. (2021); Meng et al. (2021); Xie et al. (2021) |
| Criteo | http://labs.criteo.com/downloads/download-terabyte-click-logs/ | Cerito is a dataset provided by CriteoLab, it contains one month of click logs with billions of data samples (including 39 fields). | Qu et al. (2016); Liu et al. (2018); Liu et al. (2019); Huang et al. (2019); Guo et al. (2019); Xie et al. (2019); Niu & Hou (2020); Wu et al. (2020); Zou et al. (2020b); Cheng et al. (2020); Liu et al. (2020); Zhao et al. (2020a); Khawar et al. (2020); Guo et al. (2021c) |
| Avazu | https://www.kaggle.com/c/avazu-ctr-prediction/data | The Avazu dataset was published in the Kaggle 2014 click-through rate prediction competition. The data contains 11 days data of Avazu data, ordered chronologically. In each piece of click data, there are 23 data fields such as ad id, site id, etc. | Liu et al. (2015); Chen et al. (2016b); Avila Clemenshia & Vijaya (2016); Liu et al. (2017); Qiu et al. (2018); Chan et al. (2018); Liu et al. (2018); She & Wang (2018); Liu et al. (2019); Shi et al. (2019); Guo et al. (2019); Zhang et al. (2019); Huang et al. (2019); Xie et al. (2019); Li et al. (2019); Chen et al. (2019); Chang et al. (2020); Tao et al. (2020); Niu & Hou (2020); Zou et al. (2020b); Lian & Ge (2020); Pan et al. (2020); Shi & Yang (2020); Li et al. (2020a); Li et al. (2020b); Liu et al. (2020); Livne et al. (2020); Yan et al. (2020); Song et al. (2020); Yu et al. (2020); Wang et al. (2020b); Wu et al. (2020); Zhao et al. (2020a); Zhao et al. (2020b); Zhu et al. (2020); Khawar et al. (2020); Jiang et al. (2021); Sun et al. (2021); Li et al. (2021b); Liu et al. (2021); Meng et al. (2021); Zhao et al. (2021) |
| Avito | https://www.kaggle.com/c/avito-context-ad-clicks/data | The Avito is Russia's largest user and classified commodity trading platform. Avito provided 8 datasets in the 2015 Kaggle click-through rate prediction competition, which aims to predict | Jiang et al. (2018); Ouyang et al. (2019a); Ouyang et al. (2019b); Ouyang et al. (2019c); Xie et al. (2019); Jiang et al. (2021) |
| | | whether a specific user will click on a given contextual ad. | |
| KDDCUP 2012 track 2 dataset | https://www.kaggle.com/c/kddcup2012-track2 | KDDCUP 2012 track 2 dataset gave the training instances derived from session logs of the Tencent proprietary search engine, soso.com. Each record contains 12 fields. | Wang et al. (2012); Rendle (2012b); Wang et al. (2013); Meng et al. (2016); Jiang et al. (2017); Jie-Hao et al. (2017); Ke et al. (2017); Wang et al. (2018); Wang & He (2018); Pan et al. (2019); Wang et al. (2019); Song et al. (2020); Shi & Yang (2020) |
| iPinYou | http://contest.ipinyou.com/ | The iPinYou dataset was provided by iPinYou in Global RTB (Real-Time Bidding) Bidding Algorithm Competition in 2013. This data set contains advertising bid, impression, click and conversion data, it provides effective data resources for real-time bidding (RTB) and advertising click-through rate prediction. | Zhang et al. (2014a); Liao et al. (2014); Qu et al. (2016); Shan et al. (2016); Pan et al. (2016); Zhang et al. (2016); Wu et al. (2019); Guo et al. (2019); Huang et al. (2020); Khawar et al. (2020) |
| Outbrain Click Prediction | https://www.kaggle.com/c/outbrain-click-prediction/data | The Outbrain Click Prediction dataset contains a sample of users' page views and clicks, observed on multiple publisher sites in the United States between 14-June-2016 and 28-June-2016. It is a very large relational dataset, the page views log (page_views.csv) is over 2 billion rows and 100GB uncompressed. | Yuan et al. (2020) |
| Aliyun Taobao display advertising dataset | https://tianchi.aliyun.com/dataset/dataDetail?dataId=56 | The Aliyun Taobao display advertising dataset contains click log from an online display advertising platform in Alimama Inc. There are 26 million records randomly sampled 1.1 million users for 8 days of ad display /click logs. | Li et al. (2021a); Guo et al. (2021b); |
| Huawei DIGX algorithm contest dataset 2019 | https://developer.huawei.com/consumer/cn/activity/digixActivity/Olddigixdetail/712 | The 2019 Huawei DIGX algorithm contest dataset has a total of 159,837,655 samples, and the proportion of positive samples to negative is about 1:15. The sample time is a period of six consecutive days. | Zhang et al. (2020) |
| Taobao advertising dataset | https://tianchi.aliyun.com/dataset/dataDetail?dataId=408 | The Taobao advertising dataset was gathered from the traffic logs on Taobao.com. Each ad has an ID and the associated attribute features include category ID, shop ID, brand ID and intention node ID. Other features include user features and context features such as user ID, gender, age and categorical ID of user profile. | Ouyang et al. (2021) |

| | | |
|---|---|---|
| Proprietary datasets | - | Richardson et al. (2007); Chakrabarti et al. (2008); Dembczynski et al. (2008); Cheng et al. (2012); Xiong et al. (2012); Trofimov et al. (2012); Yan et al. (2014); He et al. (2014); Zhang et al. (2014b); Ta (2015); Chen et al. (2016a); Jiang et al. (2016); Zhang et al. (2017); Huang et al. (2017); Edizel et al. (2017); Ling et al. (2017); Deng et al. (2018); Zhou et al. (2018); Pan et al. (2018); Chan et al. (2018); Zhou et al. (2019); Gligorijevic et al. (2019); Pi et al. (2019); Qu et al. (2019); Ouyang et al. (2019a); Ouyang et al. (2019b); Ouyang et al. (2019c); Gharibshah et al. (2020); An & Ren (2020); Chen et al. (2020); Pi et al. (2020); Ouyang et al. (2021); Huang et al. (2021); Zhao et al. (2021) |